\documentclass[letterpaper,12pt]{article}

\usepackage{mathptmx}

\usepackage{graphics}

\usepackage[authoryear,comma,sectionbib]{natbib}

    \usepackage{amsmath, amsthm, amsfonts, dsfont, amssymb}
    \usepackage[colorlinks=true, linkcolor=blue, urlcolor=blue, citecolor=blue]{hyperref}
    \usepackage{booktabs} 
    \usepackage{array}
    \usepackage[dvips]{graphicx}
    \usepackage{multirow}
    \usepackage{caption}
    \usepackage{setspace}
\begin{document}
\title{\LARGE Realignment in the NHL, MLB, the NFL, and the NBA}
\author{\large Brian Macdonald\footnote{\url{bmac@jhu.edu}} \qquad William Pulleyblank\footnote{ \url{william.pulleyblank@usma.edu}} \\
United States Military Academy \\  
    Department of Mathematical Sciences \\
    West Point, NY $10996$ \\
}
\date{{\footnotesize\today}}

\maketitle

\begin{abstract}
Sports leagues consist of conferences subdivided into divisions. Teams play a number of games within their divisions and fewer games against teams in different divisions and conferences.  Usually, a league structure remains stable from one season to the next. However, structures change when growth or contraction occurs, and realignment of the four major professional sports leagues in North America has occurred more than twenty-five times since $1967$.  In this paper, we describe a method for realigning sports leagues that is flexible, adaptive, and that enables construction of schedules that minimize travel while satisfying other criteria.  We do not build schedules; we develop league structures which support the subsequent construction of efficient schedules.  Our initial focus is the NHL, which has an urgent need for realignment following the recent move of the Atlanta Thrashers to Winnipeg, but our methods can be adapted to virtually any situation. We examine a variety of scenarios for the NHL, and apply our methods to the NBA, MLB, and NFL.  We find the biggest improvements for MLB and the NFL, where adopting the best solutions would reduce league travel by about $20$\%.
\end{abstract}


\tableofcontents

\section{Introduction}

The four major sports leagues in North America currently consist of thirty or thirty-two teams.  These teams are divided into divisions which are grouped to form conferences (or leagues in the case of MLB). Teams typically play the same number of home and away games against other teams in the same division, and a smaller number of games against teams in other divisions and conferences.   

The amount of travel by a team over a season is determined by three major factors:  ($1$) the distance between the team and the other cities in its division, its conference and the other conference, ($2$) the number of away games they must play against those teams, and ($3$) the scheduling of the team's away games.  The schedules are created annually and take into account a range of factors including stadium availability, holiday weekends and the possibility of making efficient road trips.  A league structure, however, will normally remain unchanged for a number of years.  Teams typically stay in the same division and conference until another team enters the league or moves to a different city.  

\begin{figure}[h!]
        \includegraphics[width=.5\textwidth]{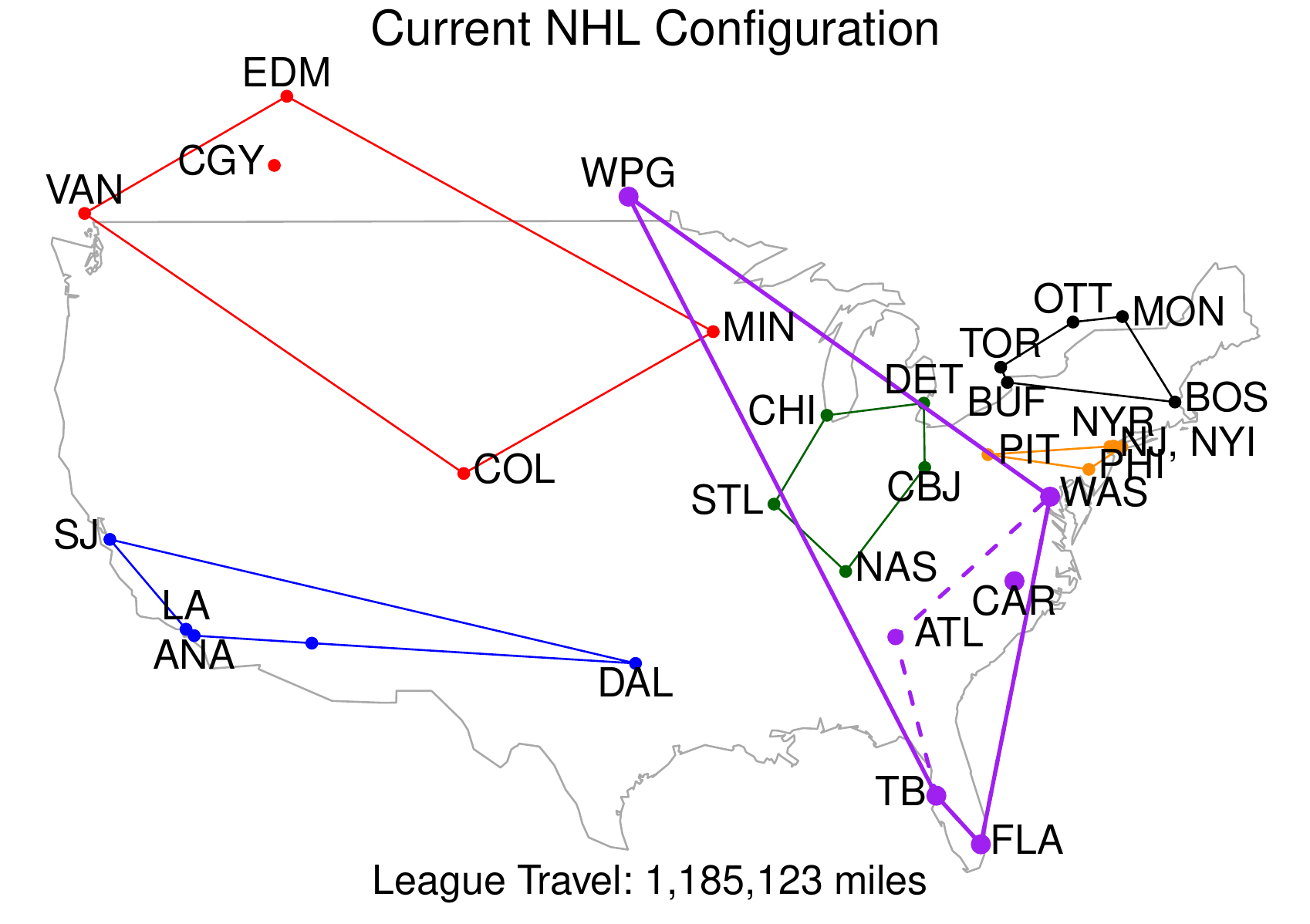}
        \includegraphics[width=.5\textwidth]{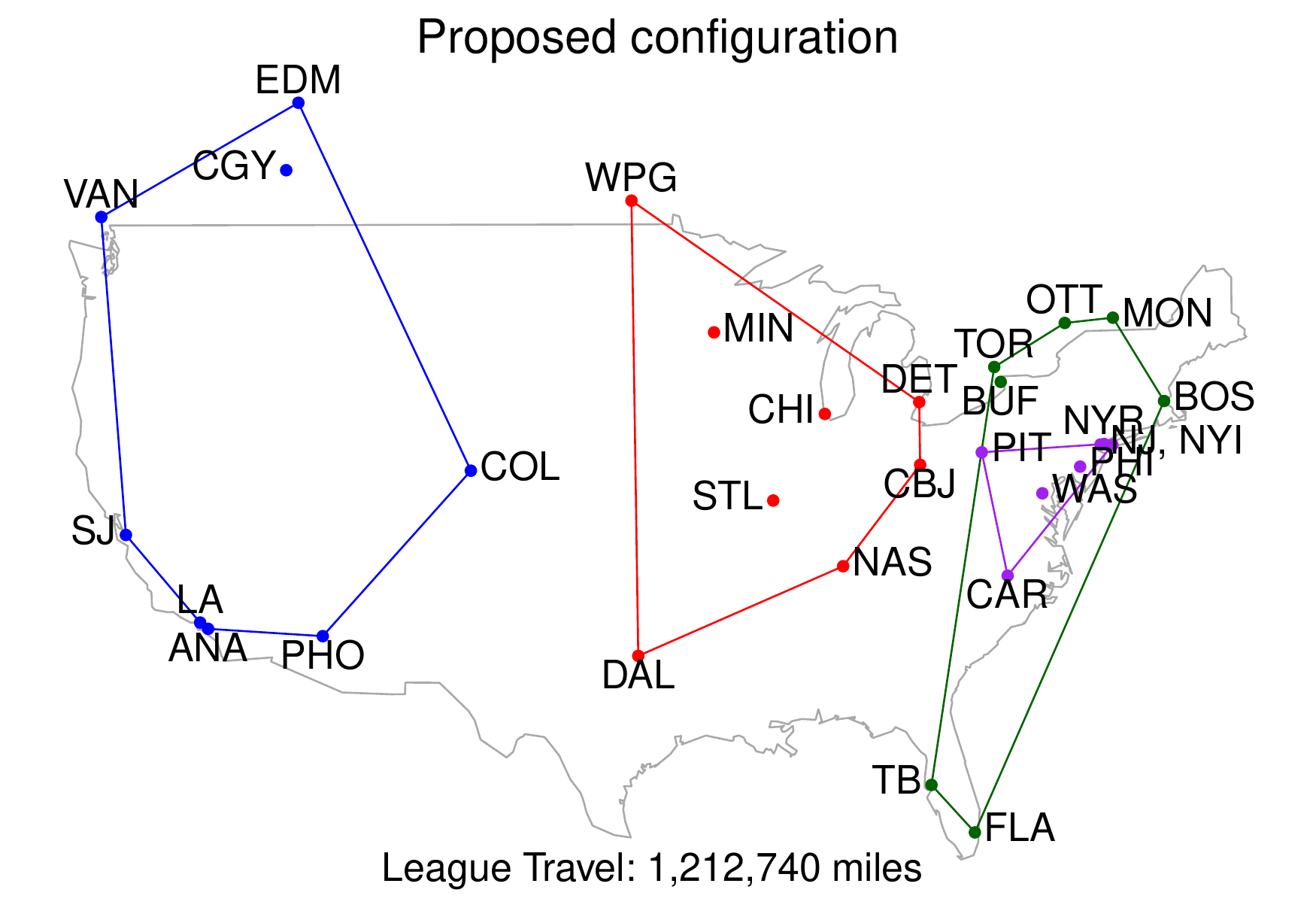}        
        \caption{(Left) The league setup before (dotted) and after (solid) the Atlanta Thrashers moved to Winnipeg.  (Right) The proposed 4-conference league structure that was approved by the NHL Board of Governors but rejected by the NHL Players' Association.}
        \label{ATL-to-WPG}
    \end{figure}

For example, in $2010$, the NHL approved the moving of the Thrashers from Atlanta (ATL) to Winnipeg (WPG), where they became the resurrected Winnipeg Jets.   In the left of Figure $\ref{ATL-to-WPG}$, we give the league setup before and after the move.  
After the move, Winnipeg remained in the Southeast division, causing significant increases in travel for Winnipeg and the other teams in that division. The NHL realized the need for realignment, and proposed the new $4$-conference configuration pictured in the right of Figure $\ref{ATL-to-WPG}$.  This proposal was subsequently rejected by the NHLPA, and a decision about realignment remains.

In this paper we focus here on optimizing the first factor mentioned above.  We develop method for structuring sports leagues that will support the construction of the most efficient annual travel schedules possible, from both the viewpoint of the league as a whole and from the viewpoints of the various teams.  Note that these viewpoints may be contradictory.  Minimizing total league travel over a season may require increasing the travel for some teams.

This league structure has important consequences for the teams. Travel costs are a major cost for teams, and major factors are the lengths and distances traveled on road trips.  Moreover, long road trips can be physically tiring for teams, which may place them at a competitive disadvantage.  

In this paper we provide the following:
\begin{itemize}
    \item{a simple ``surrogate" objective function that enables us to give an accurate estimate of total league travel incurred by a league structure without knowing the schedule;}
    \item{a fast heuristic that creates large numbers of league structures that minimize league travel and allow the inclusion of a variety of extra constraints, such as maintaining traditional rivalries and avoiding perceived inequities;} 
    \item{exact solutions to minimizing travel that show that our heuristic did succeed in constructing the optimal solution in all cases considered;}
    \item{a way to visualize these solutions, which can be helpful to humans, who ultimately make the final decision about a league's realignment plan.}
\end{itemize}

After describing our methods, we conclude by presenting results and analysis of realignment in the NHL, MLB, the NFL and the NBA.

\section{A surrogate objective function for estimating league travel}

The goal of constructing a league structure which minimizes total travel by all teams over a season faces a major problem:  the actual construction of the season schedule, which is a major factor determining travel, takes place after the league structure has been created. We deal with this by defining a surrogate measure for the goodness of a league structure which can be computed efficiently.  We compared this measure with the actual published amounts of travel by teams and found a very high correlation between this surrogate and the actual distance that each team travelled over the last several years.  

The surrogate is equal to the sum over all pairs $(i,j)$ of teams in the league of a weighted travel distance between the home cities of teams $i$ and $j$.  This is the actual distance between the cities multiplied by the number of times team $i$ plays an away game in city $j$ during the course of a season.   For example, FLA is $180$ miles away from TB, and TB plays three away games there, so the weighted distance between those two teams would be $3\times 180 = 540$. BOS is $1184$ miles from TB, and TB plays there twice, which gives $2 \times1184 = 2368$.  The ``cost" of a schedule is the sum of these weighted distances over all pairs of teams.           
            
Formally, this is defined as follows: For each pair $(i,j)$  of cities, let $d(i,j)$ denote the distance between $i$ and $j$ and let $g(i,j)$ be the number of games that team $i$ plays in $j$'s city.  These depend only on the league structure, not on the actual season game schedule. The league's weighted distance is defined as 

\begin{equation} 
	D = \sum_{(i,j)} d(i,j)\, g(i,j). \label{eq:D}
\end{equation}

\subsection{Estimating league travel} 
Minimizing the weighted distance will tend to put teams located in cities close to each other in the same division, but we would like to know if these weighted distances translate to accurate travel distances well.  We can use data from past schedules to see how well our weighted distance for a given league structure compares with actual distances traveled in previous seasons.  

We considered actual travel data for all four leagues: 
        the NHL \cite{hoag2010, hoag2011}, MLB \cite{mlb-actual-travel}, the NFL \cite{
        nfl-travel-miles-2009-1, 
        nfl-travel-miles-2010-1, 
        nfl-travel-miles-2011-1, nfl-travel-miles-2011-2, 
        nfl-travel-miles-2012-2}
        and the NBA \cite{nba-actual-travel-1, nba-actual-travel-2}. 
        The schedule, and therefore actual team travel, changes each year, so we evaluate the surrogate based on the average travel distances over several seasons. 

We find a strong linear relationship between our surrogate and actual team travel.  We describe this relationship using a linear regression model, and using the results of this model we get a predicted distance traveled for each team in the league.  
Figure $\ref{actual-vs-obj}$ shows the actual vs. the predicted travel for individual teams in the NHL over the last four years.   
\begin{figure}
\includegraphics[width=.5\textwidth]{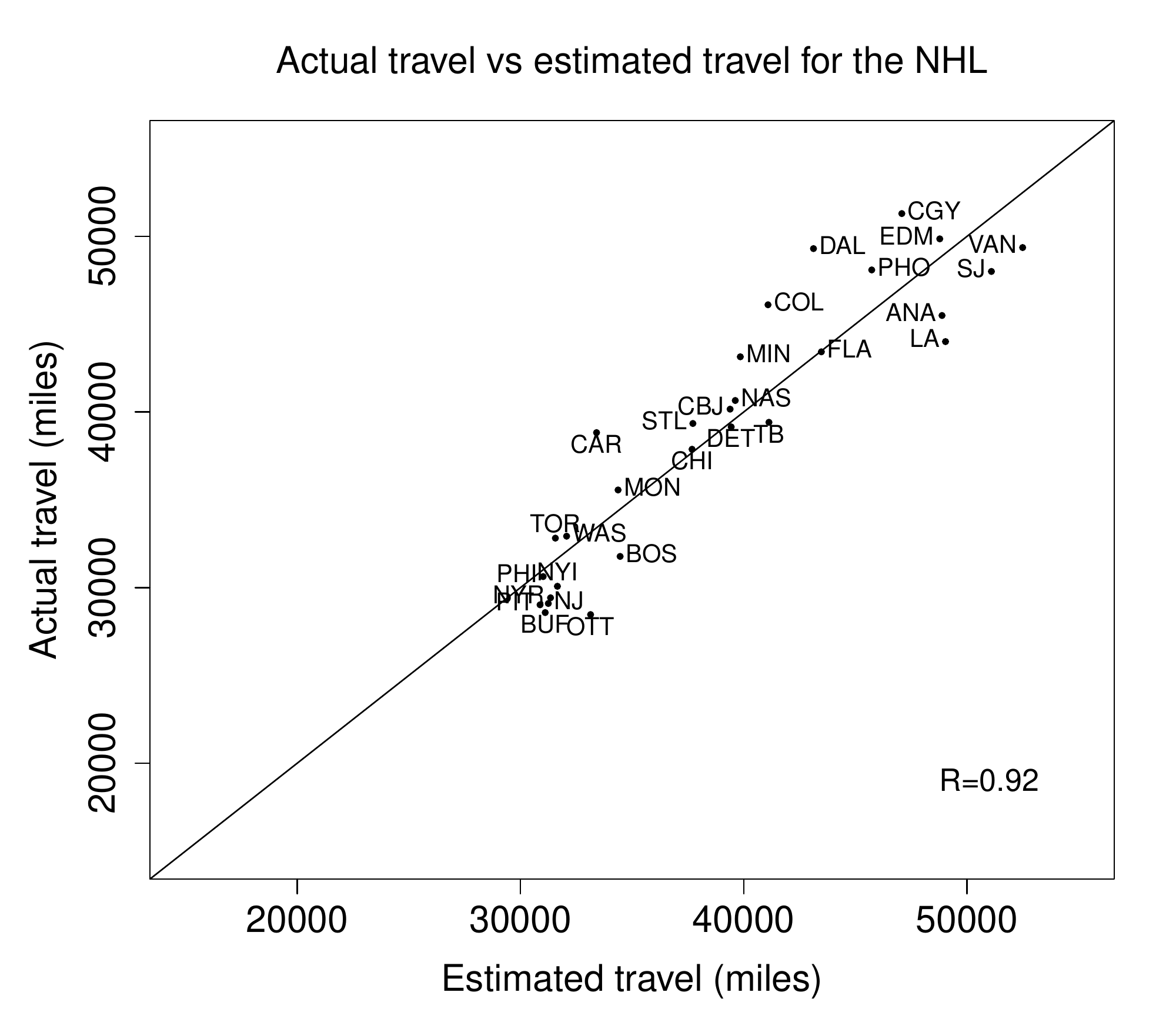}
\includegraphics[width=.5\textwidth]{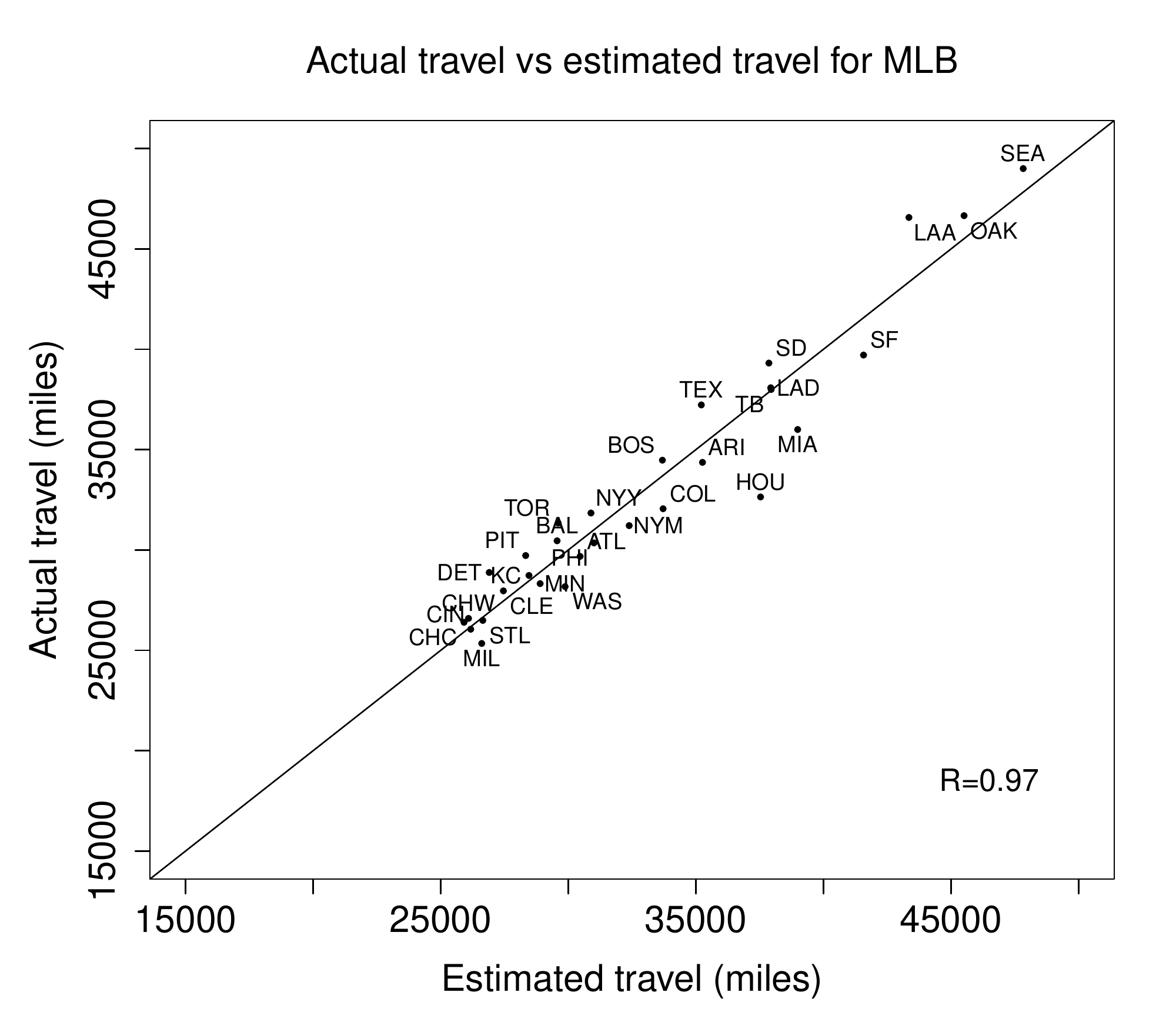}
\label{actual-vs-obj}
\caption{Actual travel versus estimated travel for the NHL (left) and MLB (right).} \label{actual-vs-obj}
\end{figure}
Our predicted team travel is very highly correlated with actual team travel.  

We also checked before ATL moved to WPG ($2008$-$09$ thru $2010$-$11$) and after ATL moved to WPG ($2011$-$12$) separately, and the fit was equally good in both cases.   In the right of Figure $\ref{actual-vs-obj}$, we give a similar plot for MLB, and again we see a very strong relationship.  We get similarly strong results for the NFL.  Our estimates for NBA team travel, while good, were not quite as strong as the others, but the NBA is the league that will be the least affected by realignment. The NBA estimates were good for most teams, but overestimated travel for the five west coast teams (LAL, LAC, SAC, GS, and POR).  

\section{A fast algorithm for generating league structures}
The \emph{convex hull} of a division is the smallest convex shape that contains the home cities of the teams in the division.  These shapes are the polygons drawn around the cities in Figure $\ref{ATL-to-WPG}$. 
Intuitively, from the standpoint of minimizing distance, it is advantageous to have the divisions be disjoint.  
We describe an efficient method for finding league structures for which the convex hulls of all divisions are disjoint.

Our algorithm first uses straight line cuts to divide the league into two disjoint conferences.  The reason that this is possible is that convex sets in the plane are disjoint only if a straight line can be drawn which separates the sets.  In addition, we can limit the separating straight lines under consideration to those lines that pass through a pair of cities.  

\begin{figure}[h!]
        \centering
        \includegraphics[width=.49\textwidth]{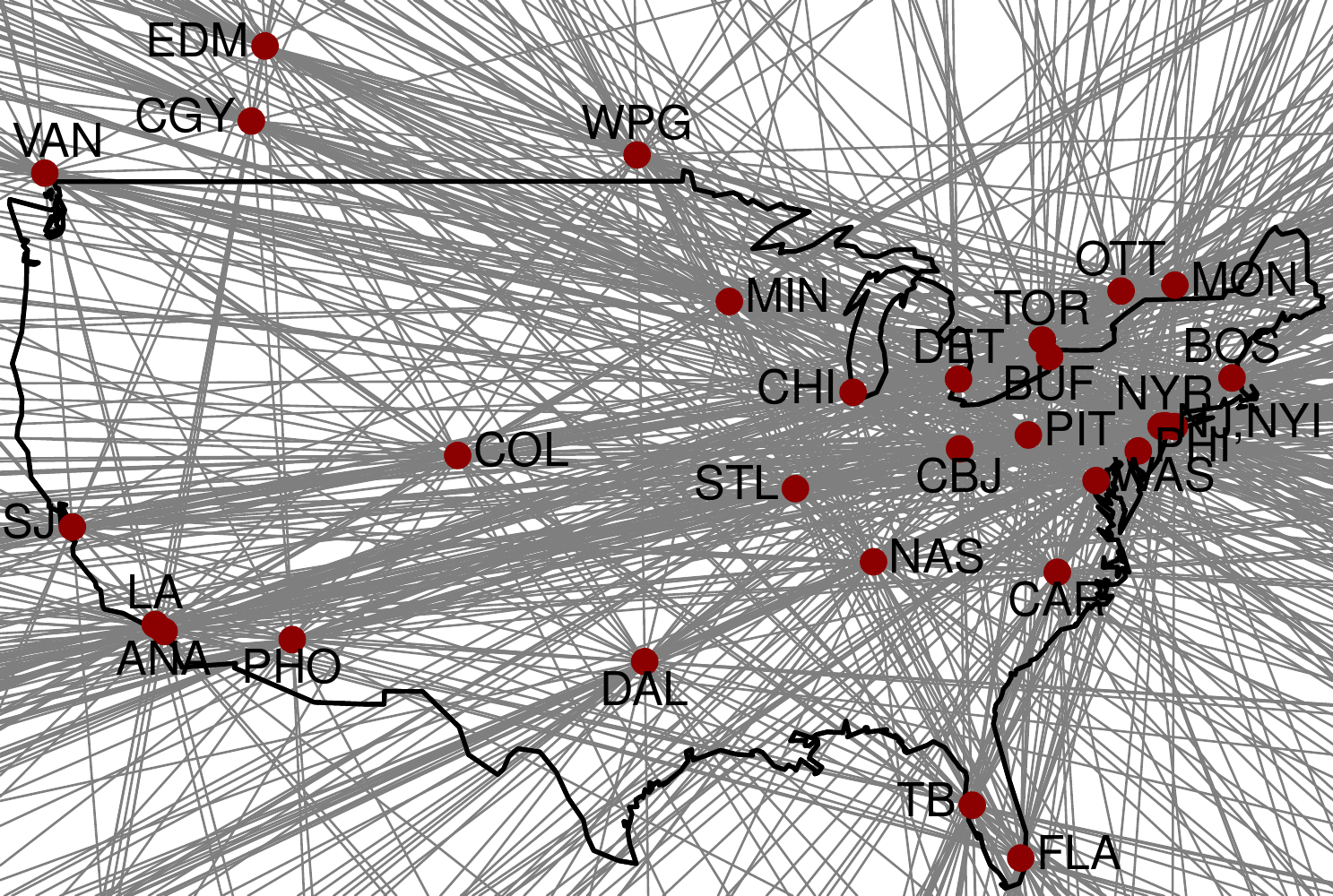}
                \includegraphics[width=.49\textwidth]{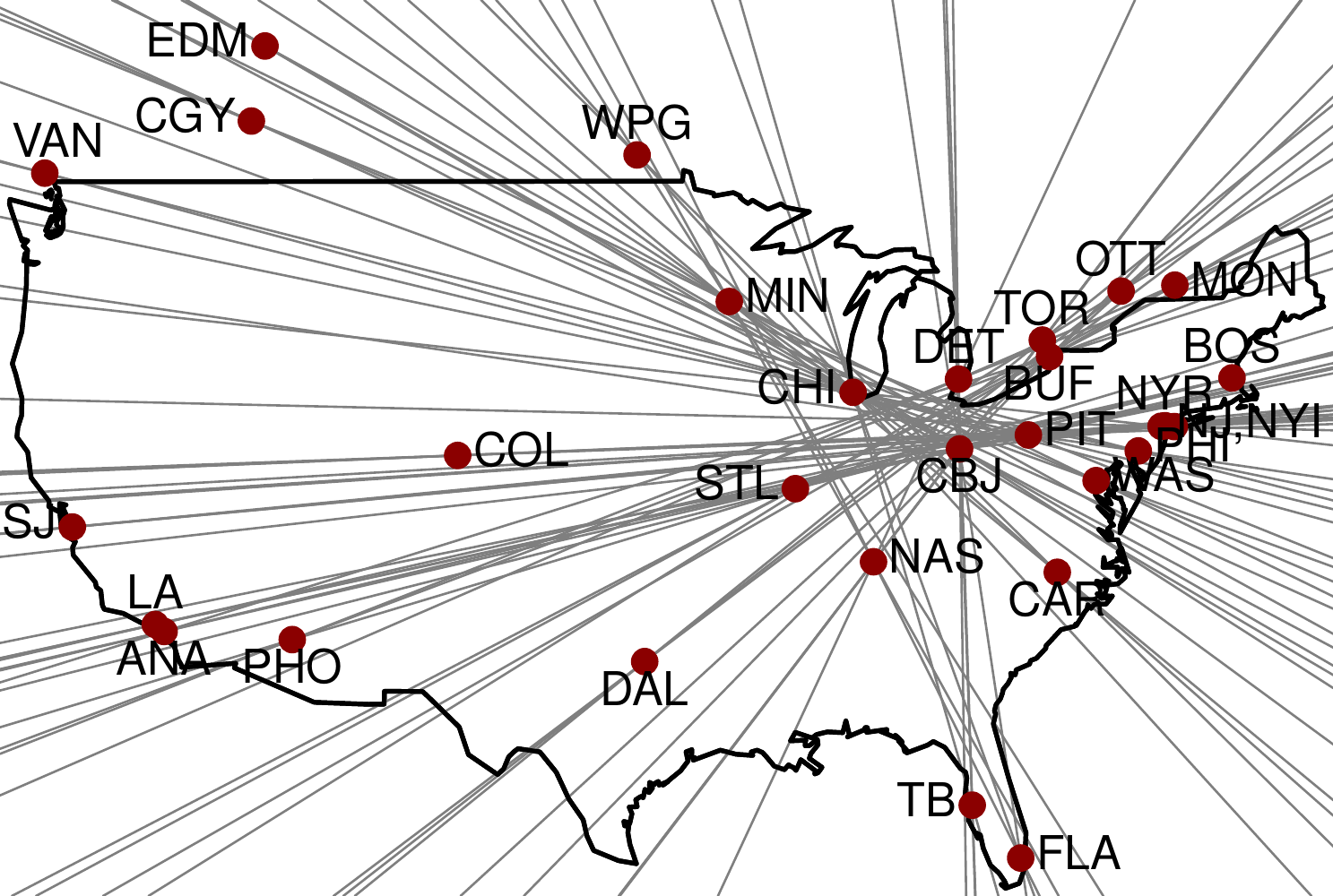}
        \caption{(Left) Lines through every pair of cities in the NHL. (Right) Lines that divide the league into equal halves.}
        \label{all-pairs}
    \end{figure}

If there are $n$ teams in the league, then the number of lines that pass through a pair of cities is ``only" $\displaystyle {n \choose 2}  = n(n-1)/2.$  In the case of a $30$ team league like the NHL, that is only $435$.  These lines are depicted in the left of Figure \ref{all-pairs}.
    
    Most such lines do not even need to be considered.  In the case of the current structure of the NHL, we only need consider separating lines that split the teams into two sets each consisting of $15$ teams to determine the conferences.  The line between LA and VAN, for example, does not split the league evenly and does not need to be considered.  We remove lines that do not split the league evenly, and we are left with the lines shown in the right of Figure \ref{all-pairs}. 

\begin{figure}[h!]
        \centering
        \includegraphics[width=.49\textwidth]{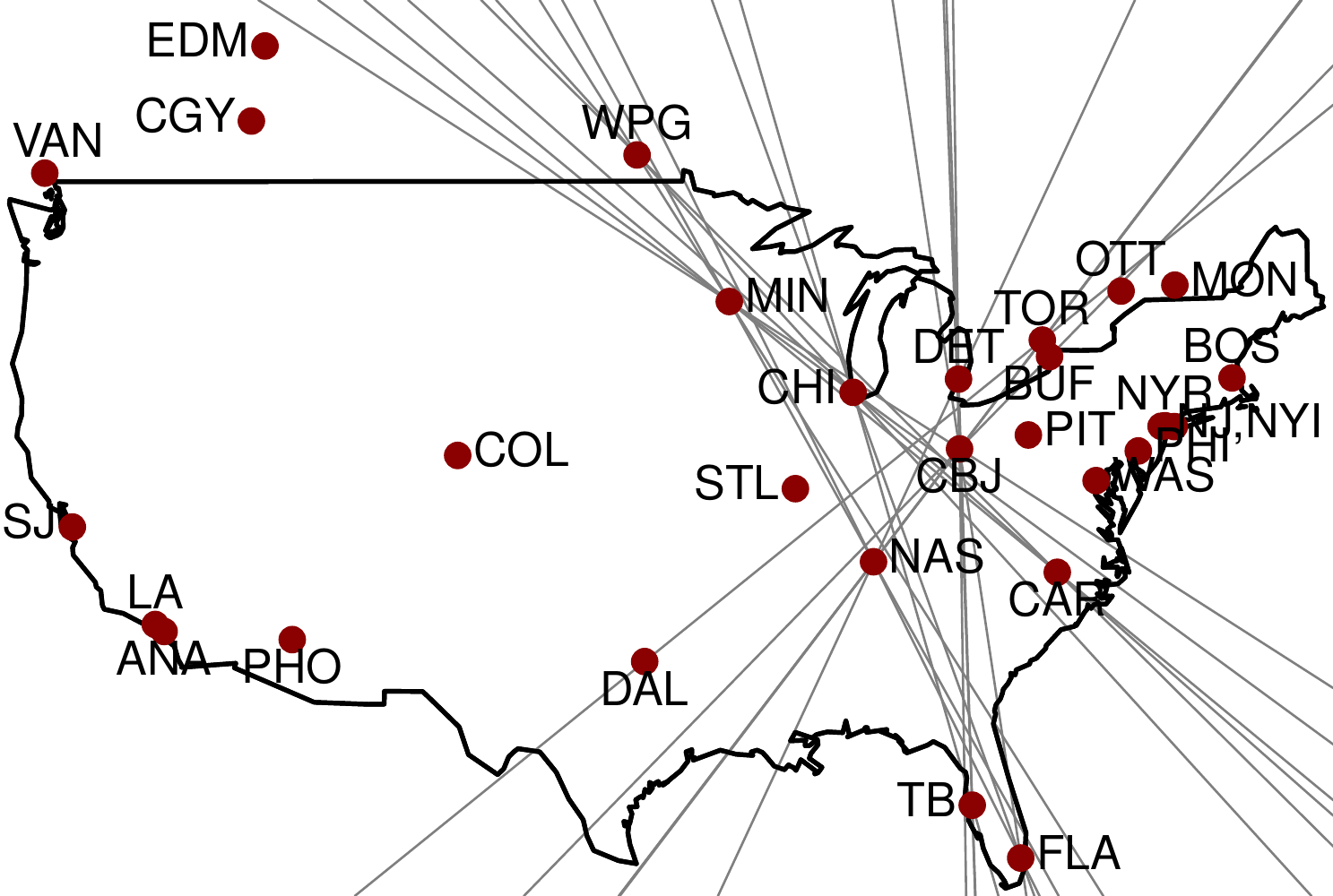}
        \includegraphics[width=.49\textwidth]{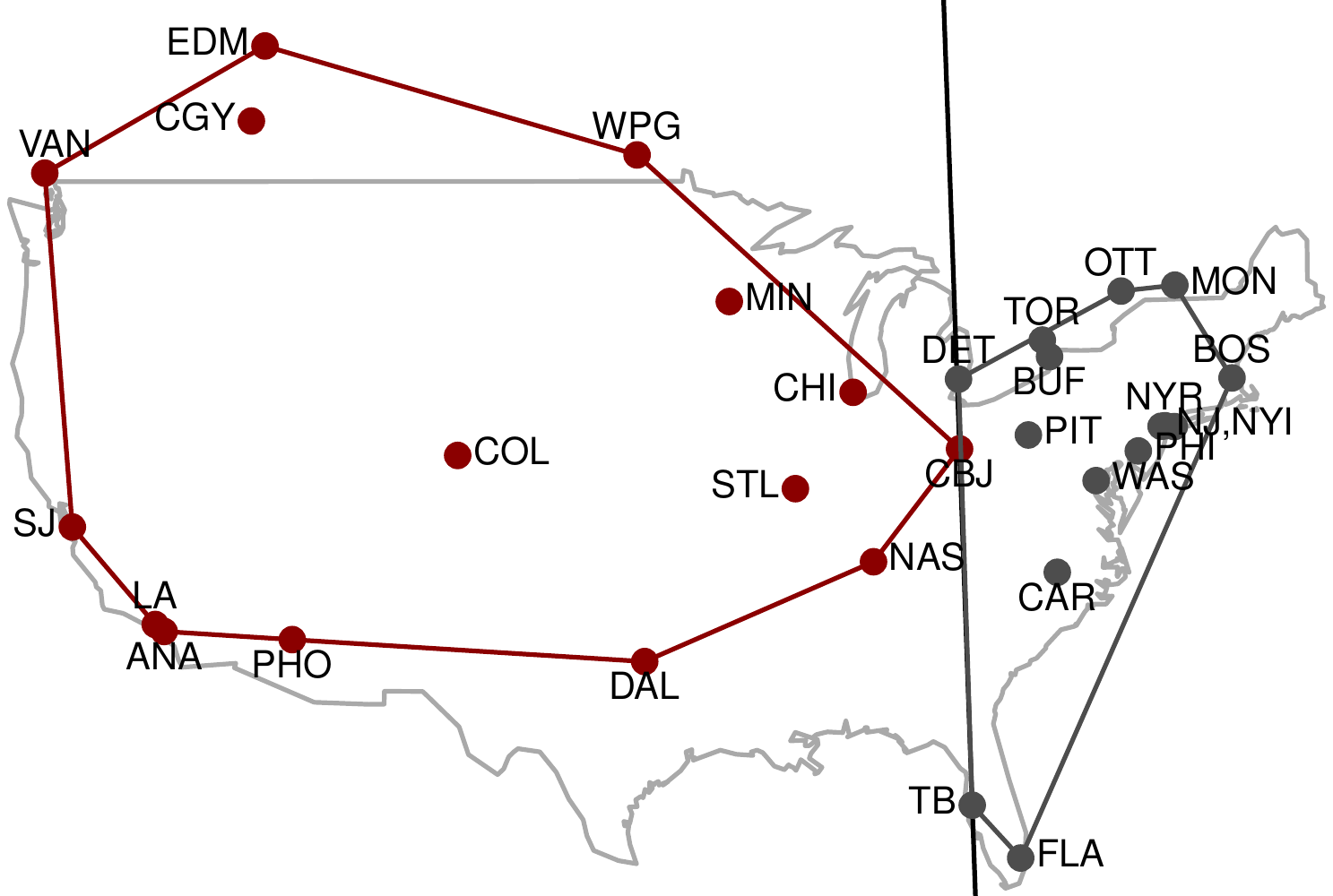}
        \caption{(Left) Lines through every pair of cities in the NHL. (Right) Lines that divide the league into equal halves.}
        \label{conf}
    \end{figure}

Some of the remaining lines would still be undesirable for splitting the league into conferences.  Lines that are too horizontal would split the league into a northern half and a southern half.  These splits would result in west coast teams being in the same conference as east coast teams, and the conferences would span all four time zones.  Even if did not care about time zones and included these lines, the resulting solutions would not be near the top of the list of the best solutions that minimize travel anyway.  So we choose to remove many of these horizontal lines. We are left with about 20 lines, which are shown in the left of Figure \ref{conf}. 
    
With each of these remaining lines, we can form two conferences, grouping the cities that are on the same side of the line into the same conference.  In the right of Figure \ref{conf} we show the convex hulls of the two conferences that result from using one of the lines.  An animation depicting the two conferences that result from using each of the remaining lines can be found at \url{www.GreaterThanPlusMinus.com/p/realignment.html}. 

We then repeat this process and split these conferences into two subgroups.  For each 15-team conference, we can find all lines that split the conference into a 10-team subgroup and a 5-team division.  The process can be repeated again for all of the 10-team subgroups, which can be split into two 5-team divisions.   The only difference in generating the divisions and generating the conferences is that we do not throw out horizontal lines while generating divisions.
    
Tens of thousands of solutions are created for each league using this method.  For example, in optimally structuring the NHL with the current league structure we generated over $100$,$000$ different candidates.   Estimated travel can be computed for the solutions we create, and we can sort the list of solutions by this estimated travel.
 
 We note that this same process can be used for the NHL, MLB, and the NBA, since each league has 30 teams split into two conferences (or leagues), each of which has three 5-team divisions.  For the NFL, which is a 32-team league, the process can be easily adapted: we find lines that split the league into two 16-team conferences, lines that split those conferences into two 8-team subgroups, and lines that split those subgroups into two 4-team divisions.
 
\subsection{Including additional constraints} 
    Our approach has another significant benefit in addition to being fast.  In restructuring a league, other factors besides travel distance are important.  There may be traditional rivalries that we want to maintain, for example, Montreal-Toronto or Pittsburgh-Philadelphia in the NHL. We may wish to sacrifice overall league distance travel in order to reduce travel for Florida and West Coast teams. We may want to keep each division within at most two time zones. Our approach deals with this very easily.  We can filter the overall set of solutions based on relevant, possibly complex, criteria and then sort based on estimated travel.  This produces several alternative structures with similar estimated travel costs, and decision-makers could choose among them using other criteria.

\subsection{Proving Optimality}  
    An important issue is how much we lose by only considering solutions generated by our algorithm. Surprisingly, it seems that we lose very little.  In the appendix we outline how truly optimal league structures can be created by solving a mixed integer programming problem (MIP).  These problems can be very difficult to solve optimally in any reasonable amount of time for situations as large as the ones we are considering.  We did however solve the MIPs corresponding to a number of the cases considered here.  In every case this established that the best solution produced by our algorithm was not just optimal among the solutions we generated, but was in fact optimal among \emph{all} solutions.

\section{Results}
\begin{figure}[h!]
        \centering
            \includegraphics[width=.9\textwidth]{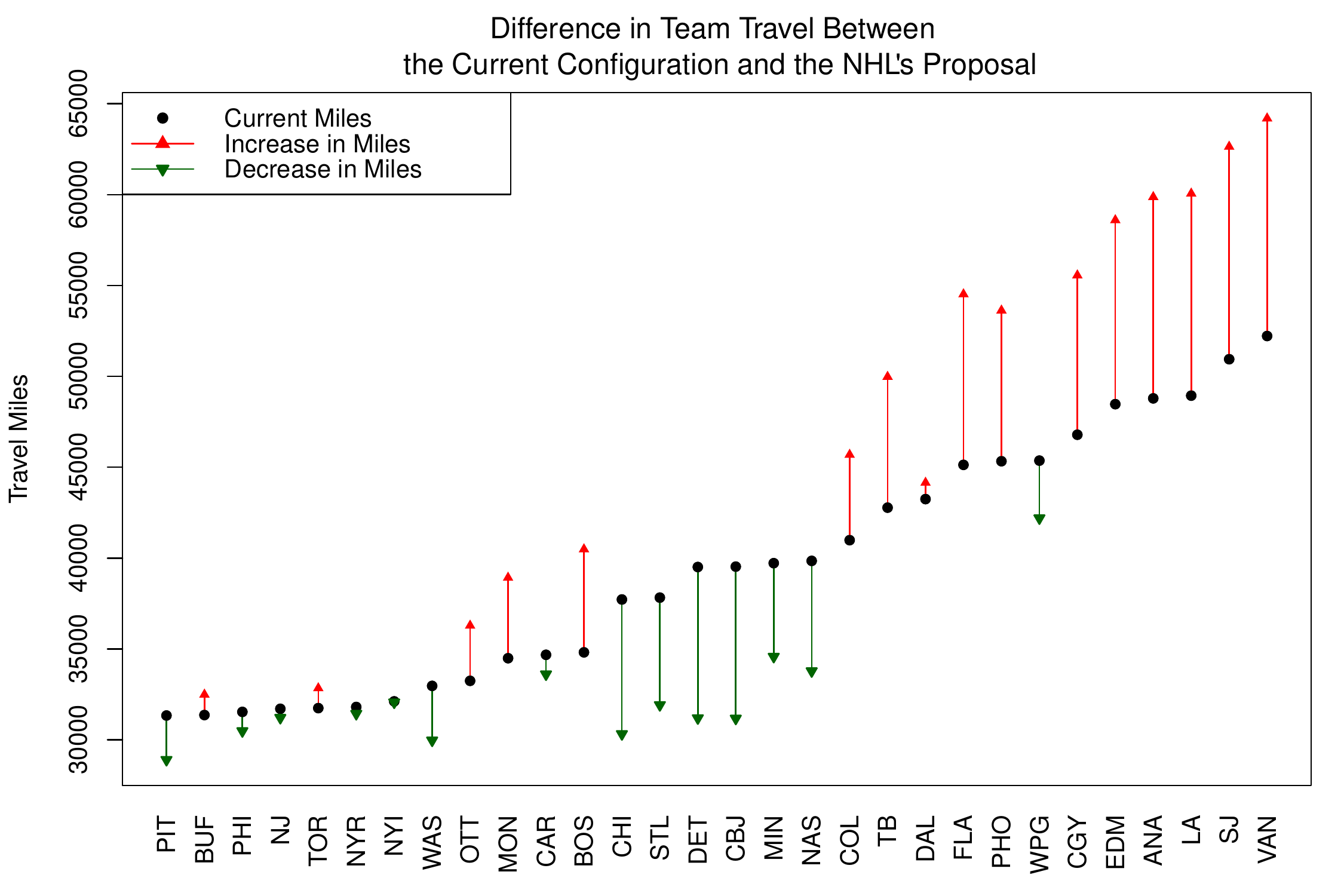}
            \caption[The difference in team travel if the NHL switches from the current configuration to the configuration proposed by the NHL Board of Governors.]{The difference in team travel if the NHL switches from the current configuration to the configuration proposed by the NHL Board of Governors.  Black dots indicate current travel for each team.  Red indicates that a team would have worse travel (more miles) in the NHL's proposed configuration.  Green indicates that a team would have better travel (fewer miles) in the proposed configuration.  
            }
            \label{current-vs-proposed}
        \end{figure}

Figure $\ref{current-vs-proposed}$ shows the estimated difference in team travel if the NHL switched from the current configuration to the configuration proposed by the NHL Board of Governors, which were depicted in Figure $\ref{ATL-to-WPG}$.  
Note that most teams would have more travel, including the teams that already have the worst travel (the west coast and Florida teams near the top).

\subsection{NHL Realignment}         
\begin{figure}[h!]
                        \includegraphics[width=.5\textwidth]{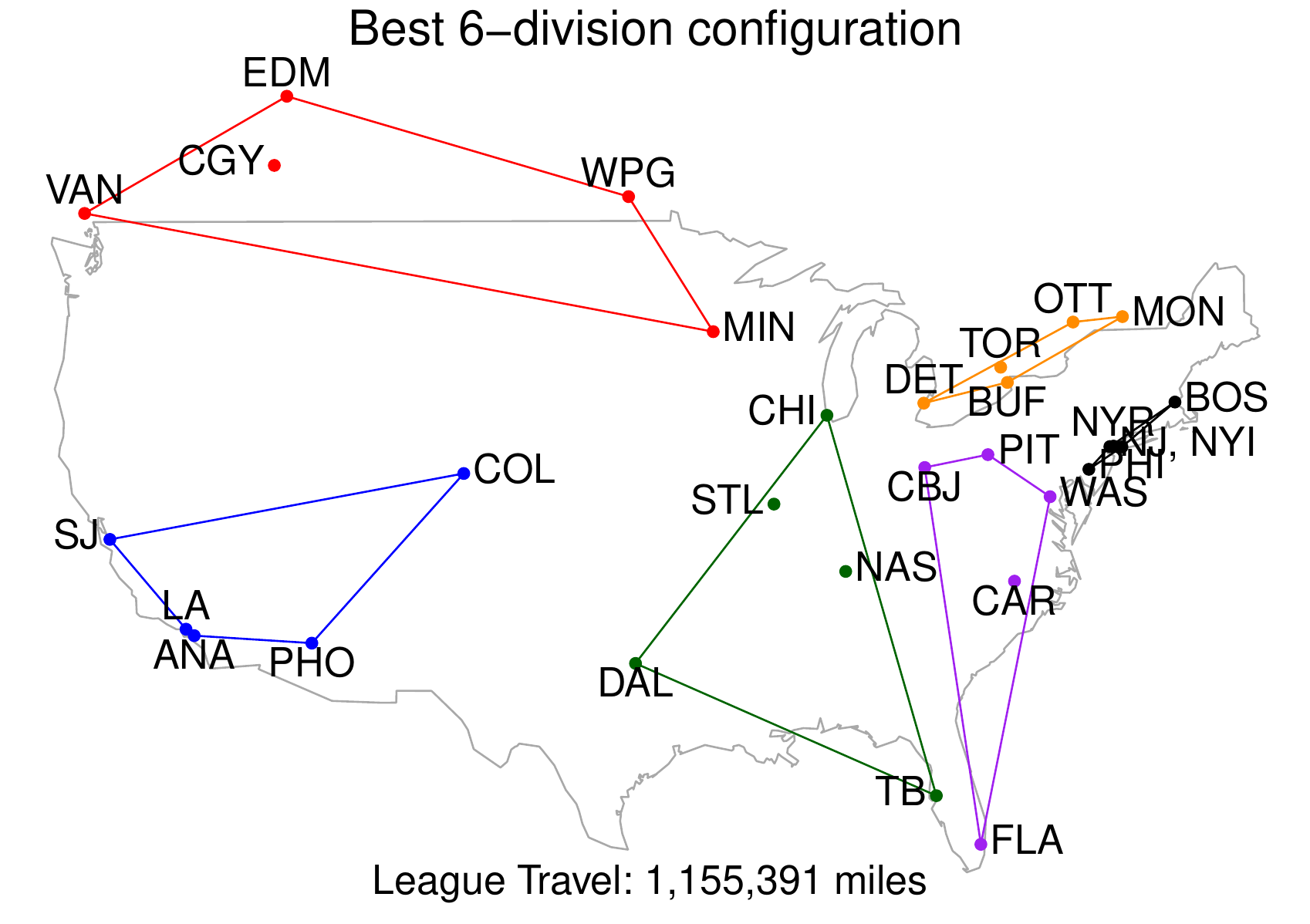}
                        \includegraphics[width=.5\textwidth]{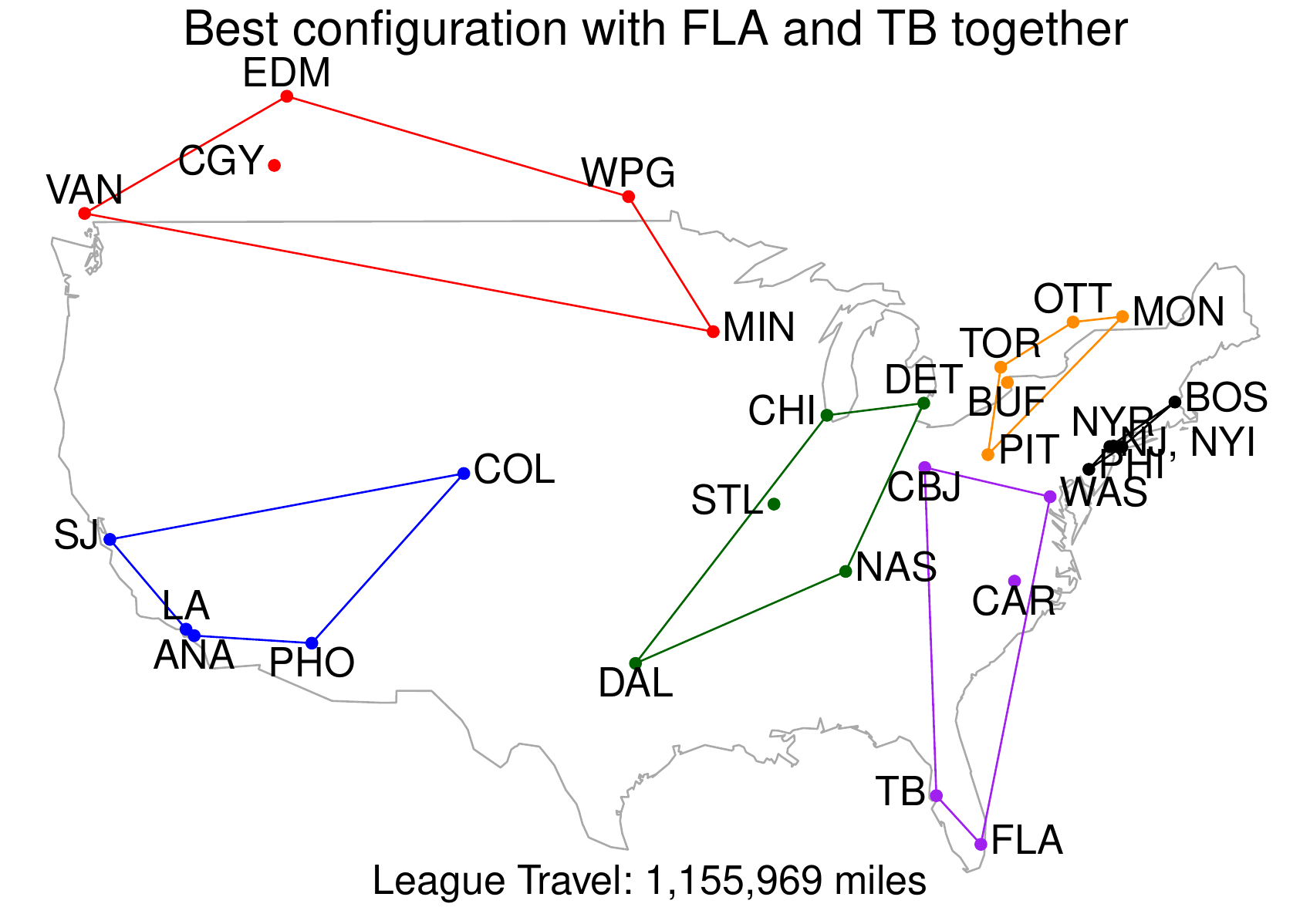}
                        \caption{(Left) The best 6-division configuration.  (Right) The best  with TB and FLA together.}
                        \label{with-and-without-constraints}
                    \end{figure}

        We now give results for the best configurations in the NHL under a variety of constraints.
        In the left of Figure $\ref{with-and-without-constraints}$, we give the best league configuration for the current structure of $6$ divisions of $5$ teams each.  Interestingly, in this case, Florida (FLA) and Tampa Bay (TB) are not in the same division.  In fact, they are not even in the same conference, as TB moves west and both Detroit and Columbus move east.  So while this is the configuration that minimizes total league distance, it would probably be undesirable based on other factors.  Ideally, we could minimize league distance, but also minimize distance traveled by the teams that have it the worst, the west coast and Florida teams.
        
        Fortunately, we can easily add constraints to the problem.  For example, we can allow only solutions in which FLA and TB are in the same division.  In the right of Figure $\ref{with-and-without-constraints}$, we give the best solution subject to this constraint.
                Note this solution would only cost the league a few hundred travel miles, a small price to pay for keeping those teams together.  This solution also minimizes travel for the west coast and Florida teams. 

    \begin{figure}[h!]
    \centering
     \includegraphics[width=.8\textwidth]{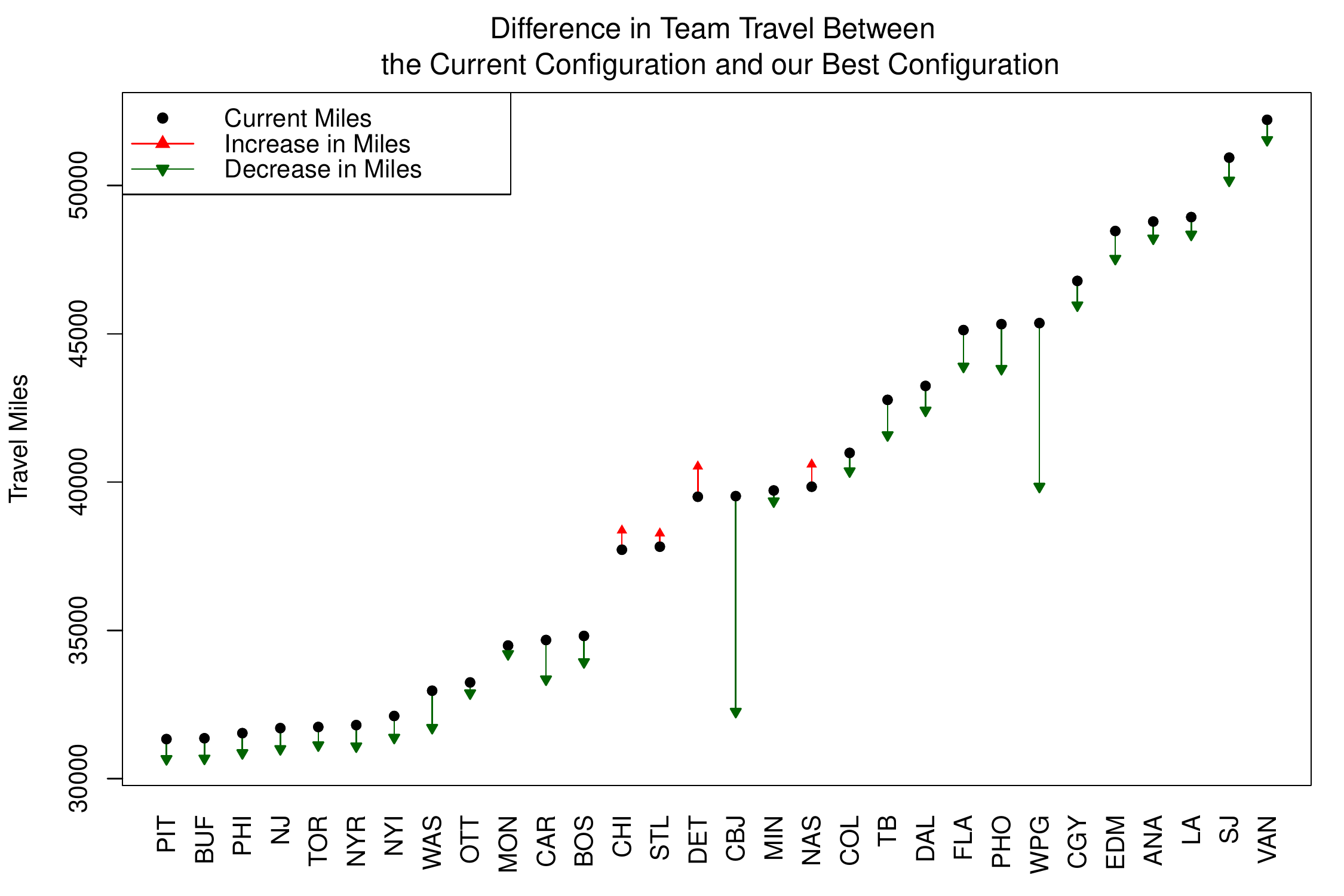}

        \caption{The difference in team travel if the NHL switches from the current configuration to our best $6$-division solution. }
        \label{current-vs-best}
    \end{figure}
%
In Figure $\ref{current-vs-best}$, we give the difference in team travel if the NHL switches from the current configuration to our best $6$-division solution. Most teams would have better travel with our solution, including the teams that have the worst travel (the west coast and Florida teams near the top). Not surprisingly, Winnipeg's travel would improve significantly.  Columbus would also have much better travel because they would replace Winnipeg the Southeast Division and join the Eastern Conference.    

\begin{figure}[h!]
    \centering
        \includegraphics[width=.8\textwidth]{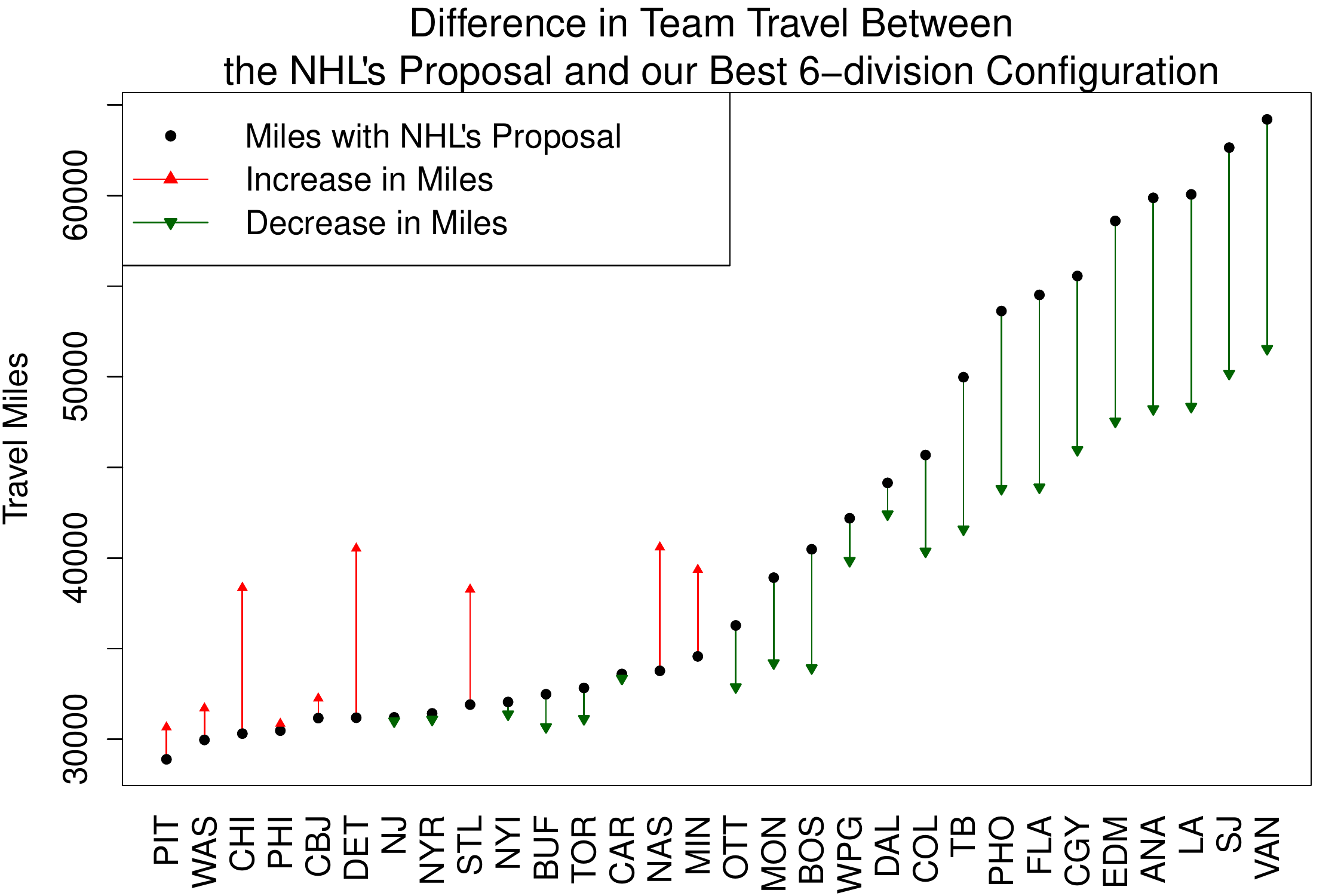}
        \caption{The difference in team travel between using the NHL's proposed configuration and our best $6$-division configuration.  
        }
        \label{proposed-vs-best}
    \end{figure}

In Figure $\ref{proposed-vs-best}$, we give the difference in team travel between using the NHL's proposed configuration and our best $6$-division configuration.   Most teams would have  significantly better travel with our solution compared to the NHL's proposal.  In particular, the west coast and Florida teams would have significantly better travel with our configuration, which can be seen in the upper right of Figure $\ref{proposed-vs-best}$.
        
\begin{figure}[h!]
                \includegraphics[width=.5\textwidth]{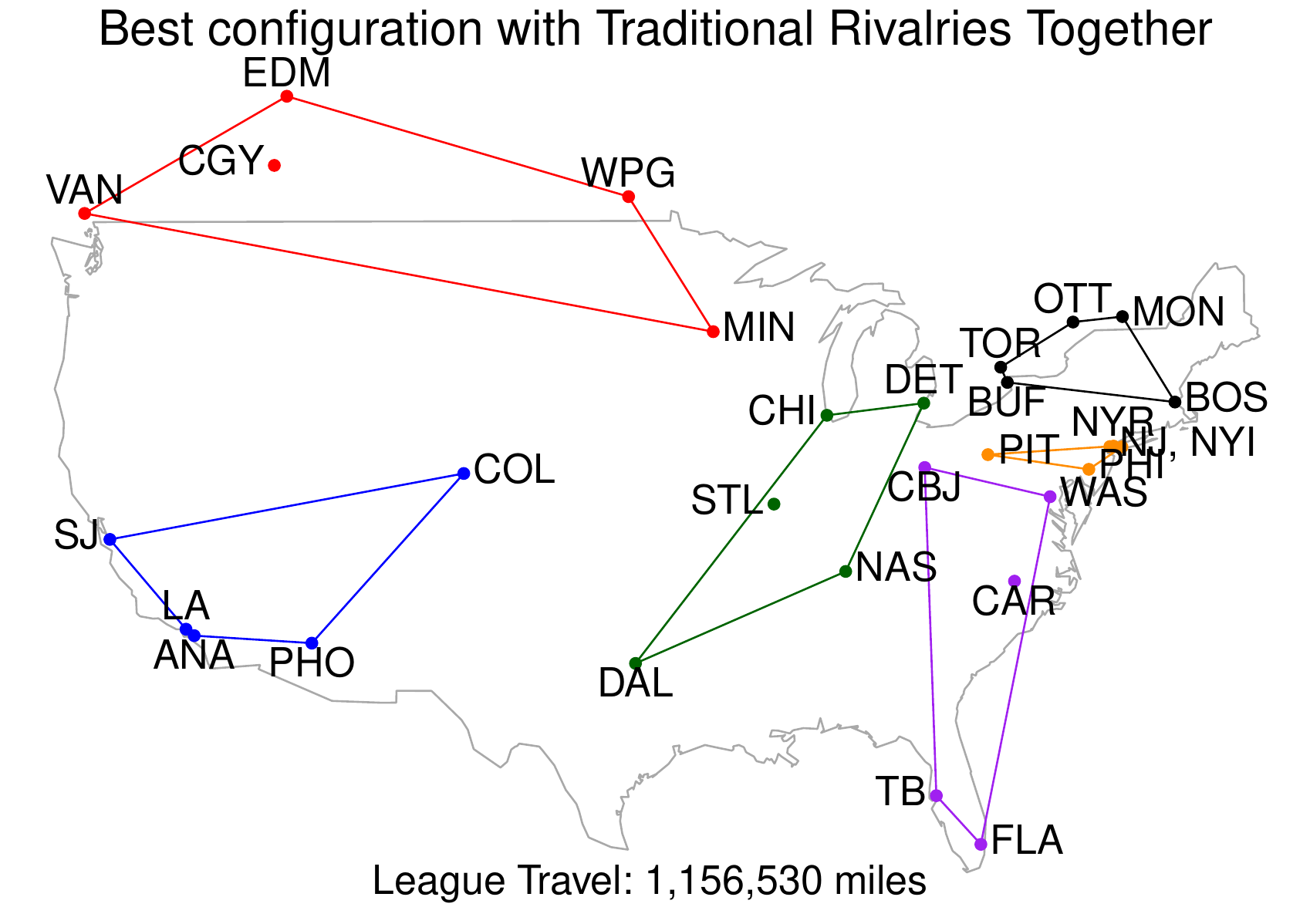}
                \includegraphics[width=.5\textwidth]{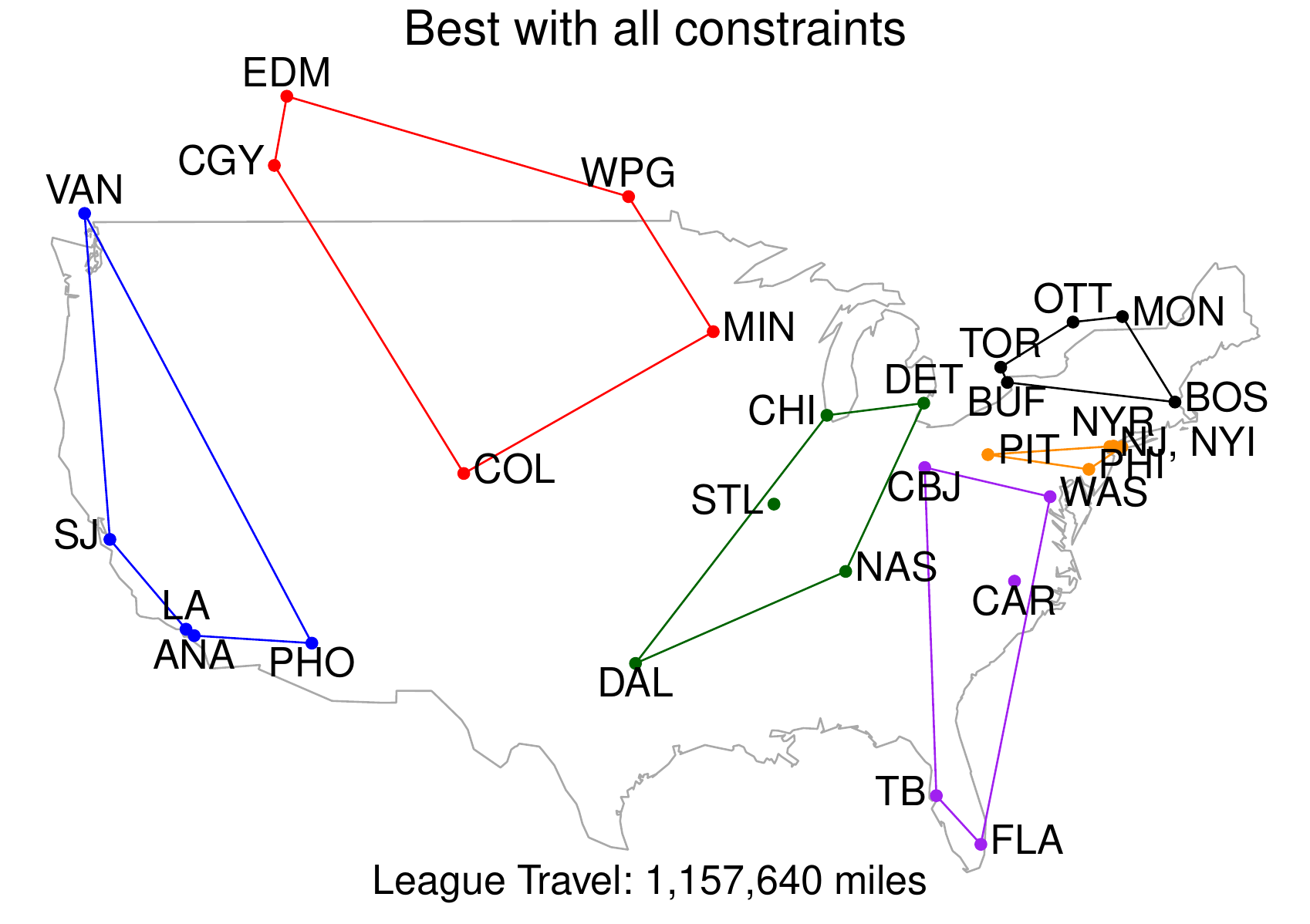}
                \caption{(Left) The best configuration with rivals together.  (Right) The best configuration with two additional constraints: at most $3$ Canadian teams can be in one division, and a division can span at most two different times zones.  }
                \label{with-more-constraints}
            \end{figure}

        There may be other constraints that one would like to add.  For example, note that in Figure $\ref{with-and-without-constraints}$, Philadelphia (PHI) and Pittsburgh (PIT) are in different divisions.  
        The NHL may prefer that PHI and PIT remain together, and that other traditional rivals remain in the same division.  In the left of Figure $\ref{with-more-constraints}$, we give the best solution that keeps the following teams together: TB and FLA; PHI and PIT; NY Rangers, NY Islanders, and NJ Devils; Calgary and Edmonton; Anaheim and Los Angeles.  
        Although this configuration is not the ``best'' according to distance, the league would travel only about $1$,$000$ miles more in this case, and the benefits of keeping these rivals together would likely far outweigh the costs of a minimal increase in travel miles.  Also, this solution still minimizes travel for west coast and Florida teams.

        Note that so far, the two western most divisions have been the same.  
        In fact, most of the top $100$ solutions have this configuration out west.  
        Note however that Minnesota is with four Canadian teams.  It has been reported that the NHL may try to avoid four Canadian teams in the same division with one lone American team.  Also, note that this division crosses three time zones, another undesirable property.  
        
        Fortunately, we can easily add yet another constraint to our problem, and require that at most $3$ Canadian teams be in one division.  The best solution in this case is given in the right of Figure $\ref{with-more-constraints}$.  This solution costs the league only about $2$,$000$ miles more than the optimal solution, and costs west coast teams an additional $3$,$400$ miles.  We give some summary statistics for this configuration, and many other configurations, in Table $\ref{summary-table}$.

\begin{figure}[h!]
                    \includegraphics[width=.5\textwidth]{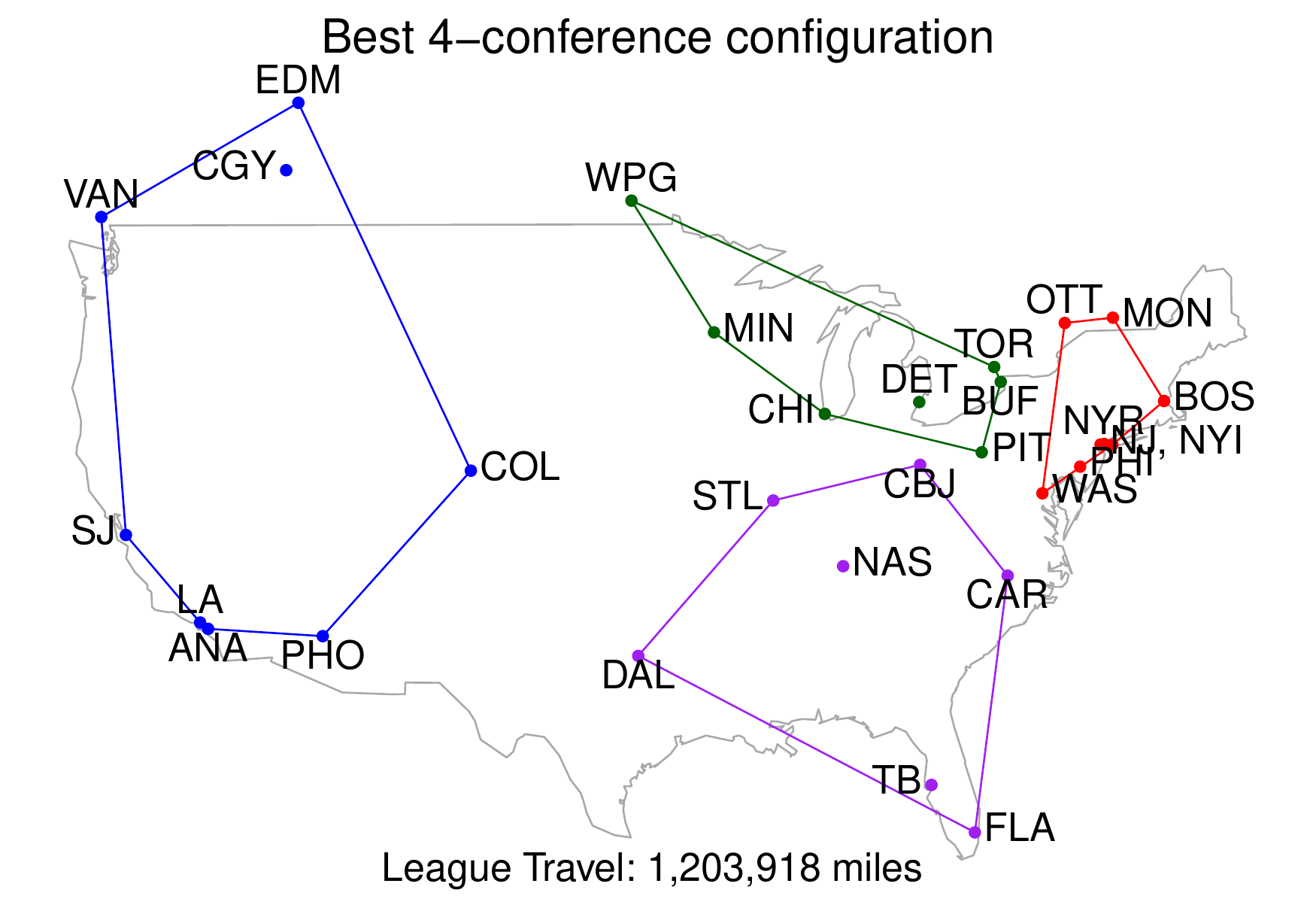}
                    \includegraphics[width=.5\textwidth]{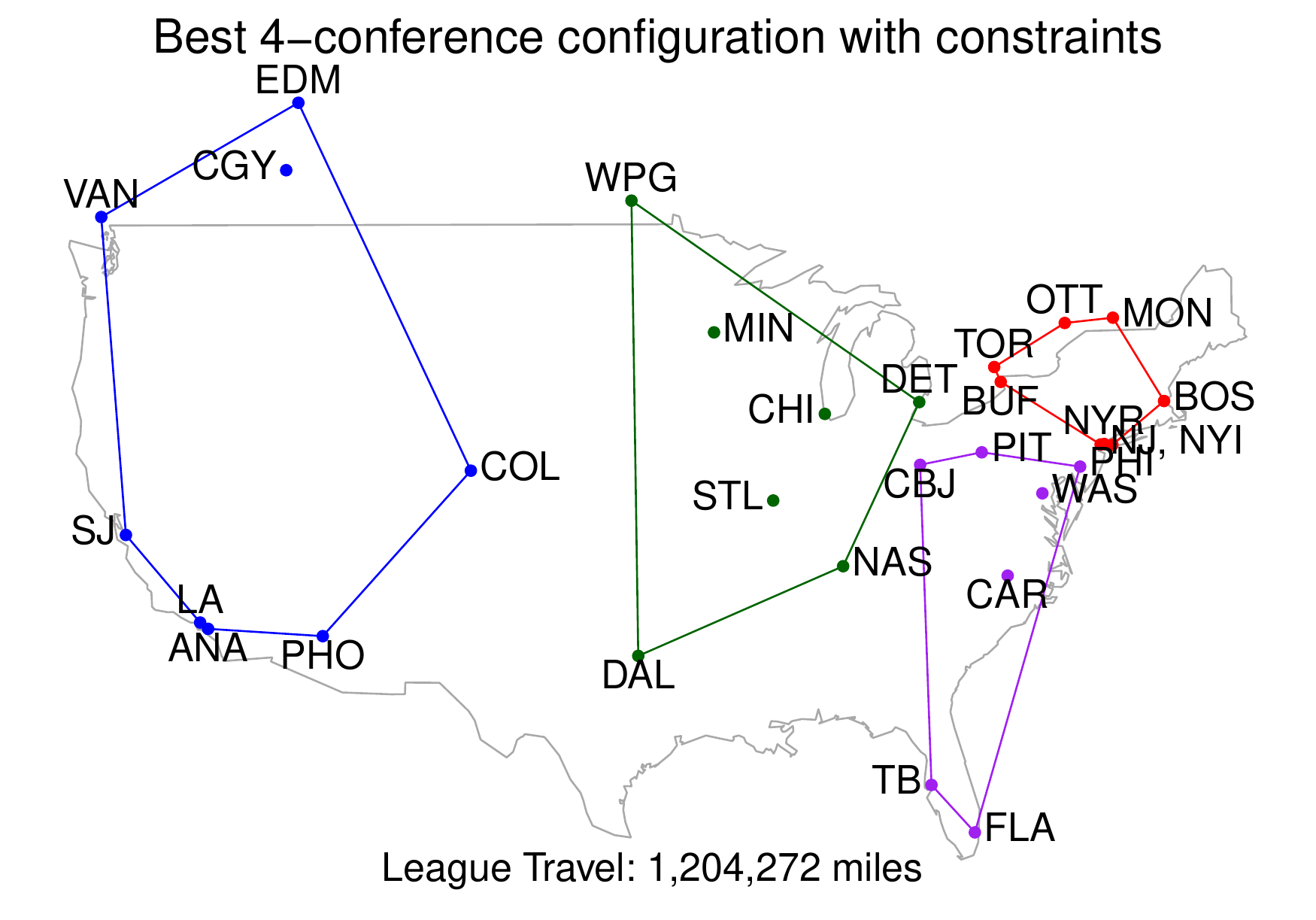}
                    \caption{(Left) The proposed $4$-conference structure.  (Right) The ``best'' configuration for a $4$-conference structure, that satisfies the same additional constraints as before.}
                    \label{4-conferences}
                \end{figure}

    \paragraph*{Best $4$-conference configurations}
        Our methods are not restricted to the current setup of $6$ divisions and $5$ teams in each division.  Since the NHL recently proposed a $4$-conference structure (see Figure $\ref{ATL-to-WPG}$), we give the best $4$-conference structure in Figure $\ref{4-conferences}$.  Also, we give the best solution using the same additional constraints as before in the right of Figure $\ref{4-conferences}$.
        We note that the proposed $4$-conference structure is typically much more costly than the current $6$-division structure.  The best $4$-conference solutions are worse than the best $6$-division solutions, resulting in increase of about 85,000 miles.  Much of this difference is due to the more balanced schedule that was proposed along with this realignment.
        
\begin{figure}[h!]
            \centering
                \includegraphics[width=.7\textwidth]{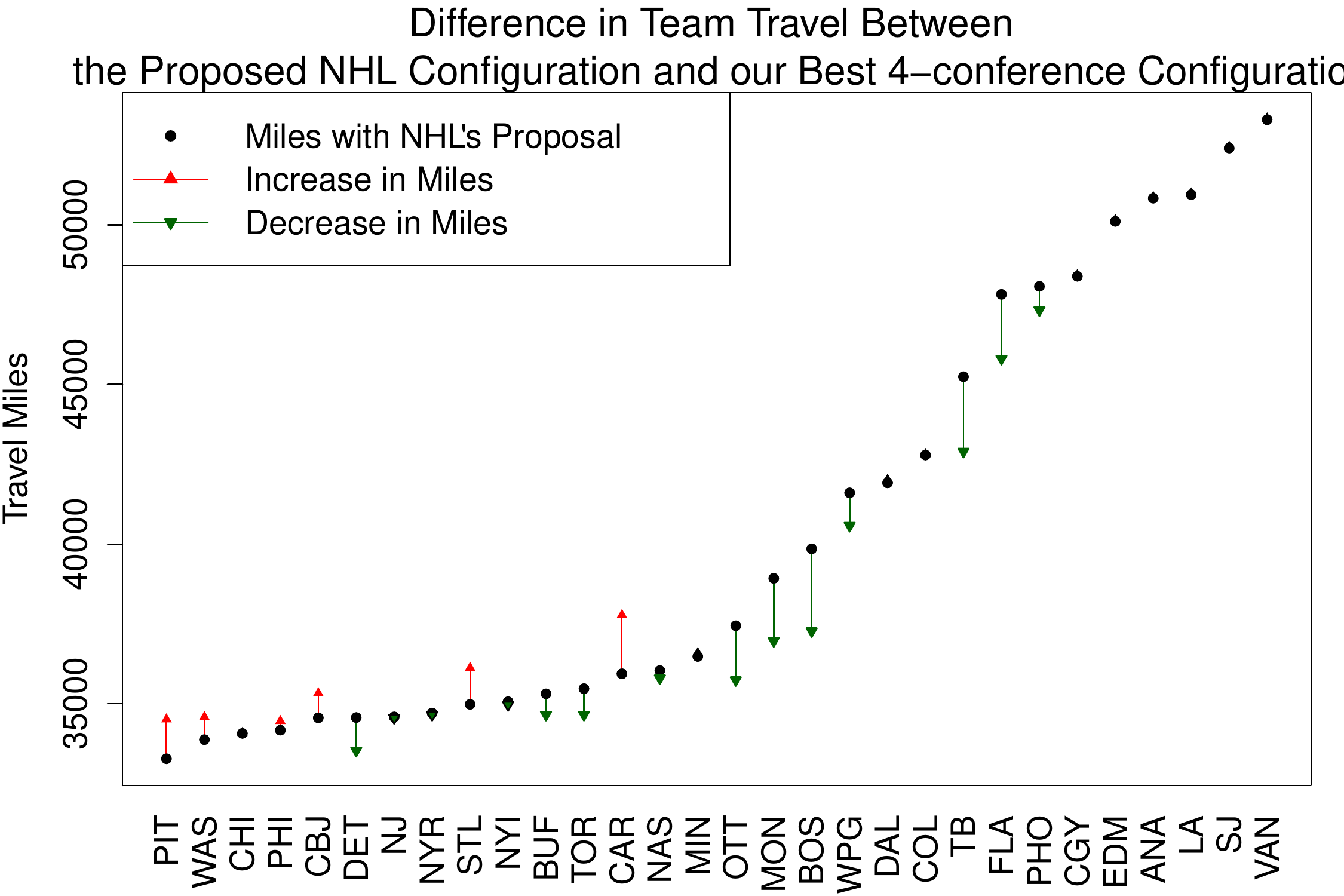}
                \caption{The difference in team travel between using the NHL's proposed configuration and our best $4$-conference configuration.  
                }
                \label{proposed-vs-best-4-conf}
            \end{figure}

\begin{figure}[h!]
        \includegraphics[width=.5\textwidth]{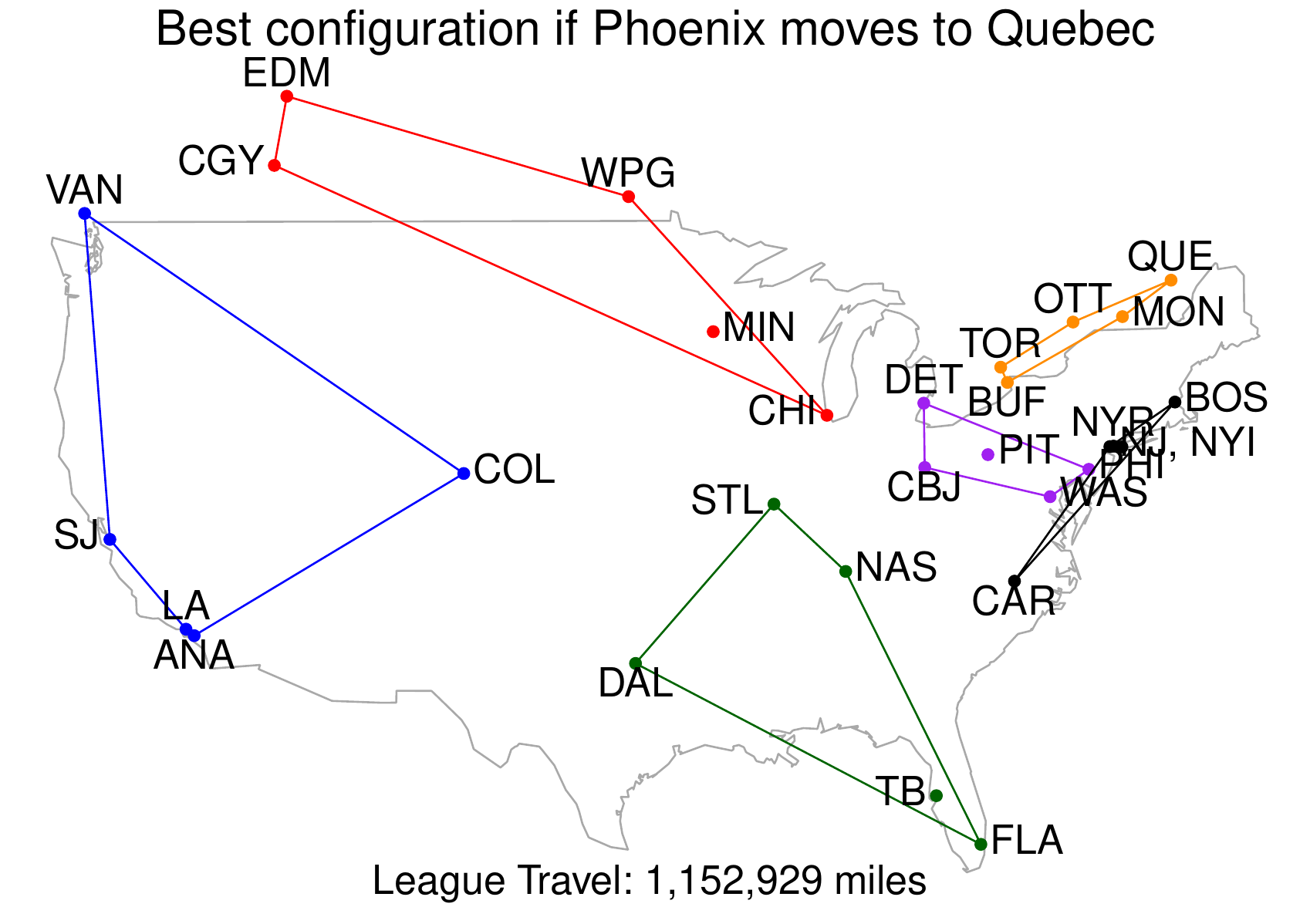}
        \includegraphics[width=.5\textwidth]{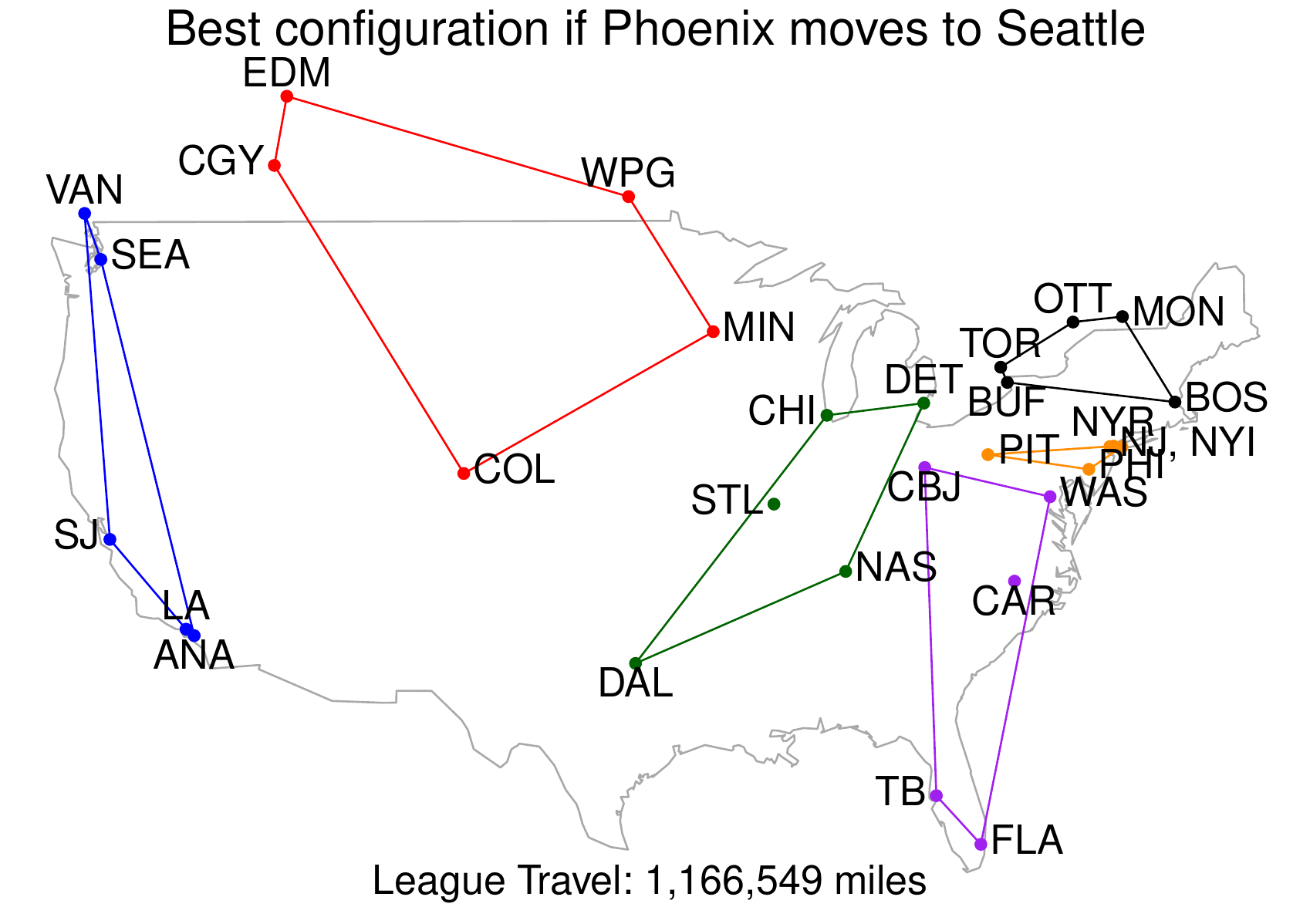}
        \includegraphics[width=.5\textwidth]{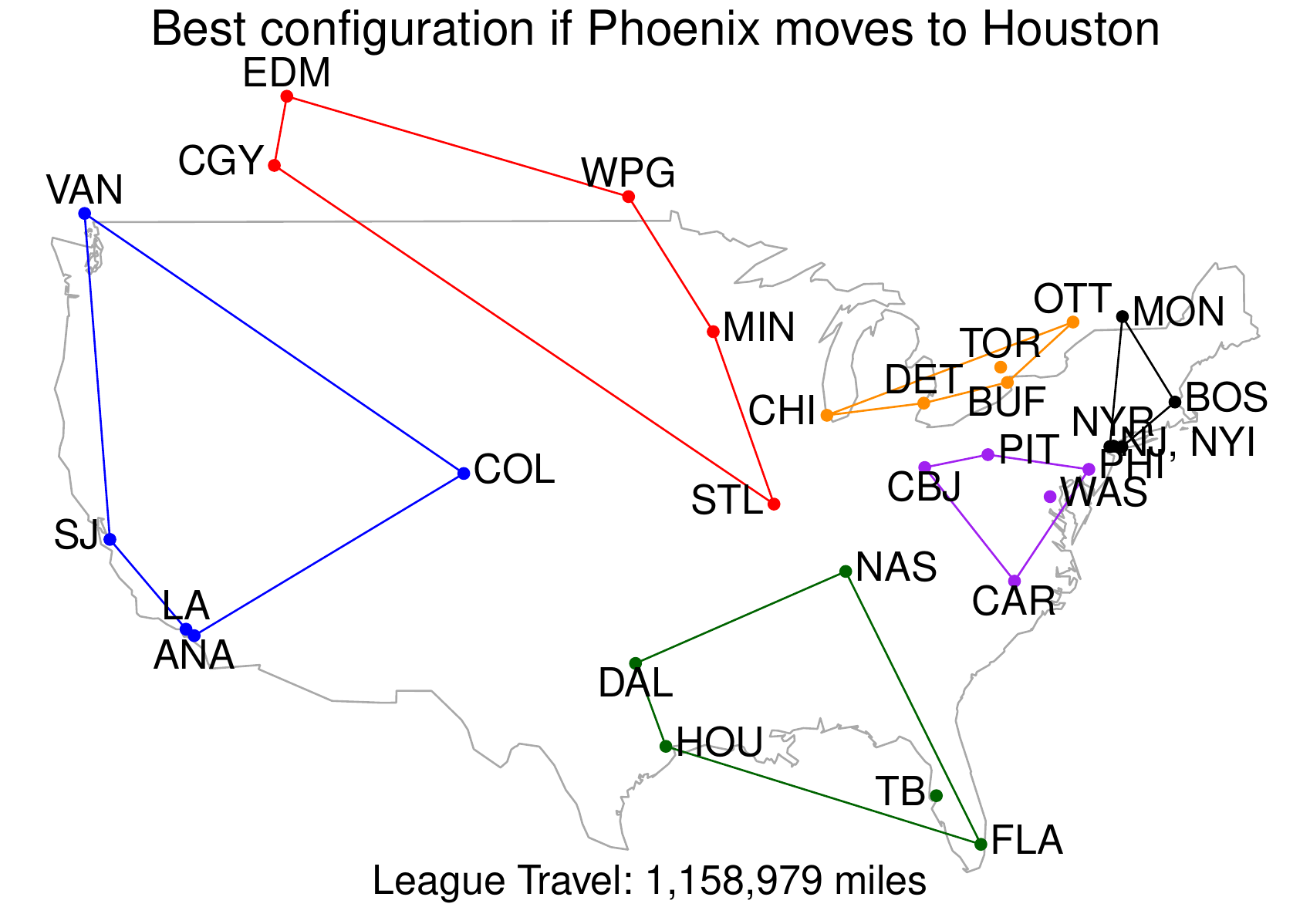}
        \includegraphics[width=.5\textwidth]{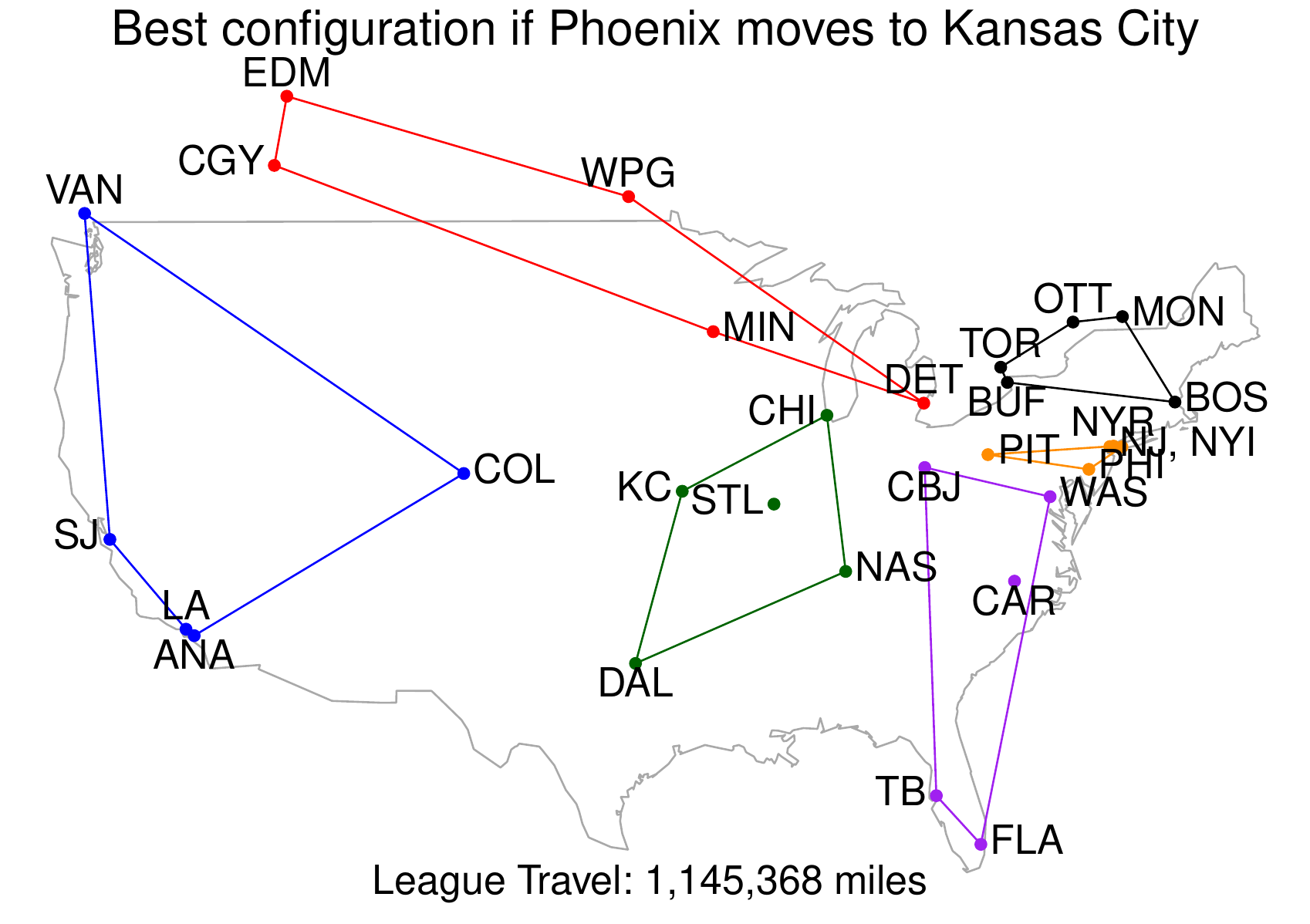}
        \caption{The best solution if PHO moves to QUE (top left), SEA (top right), HOU (bottom left), and KC (bottom right).  In all cases, we have used the same additional constraints as before. }
        \label{moves}
    \end{figure}

\begin{figure}[h!]
                \includegraphics[width=.5\textwidth]{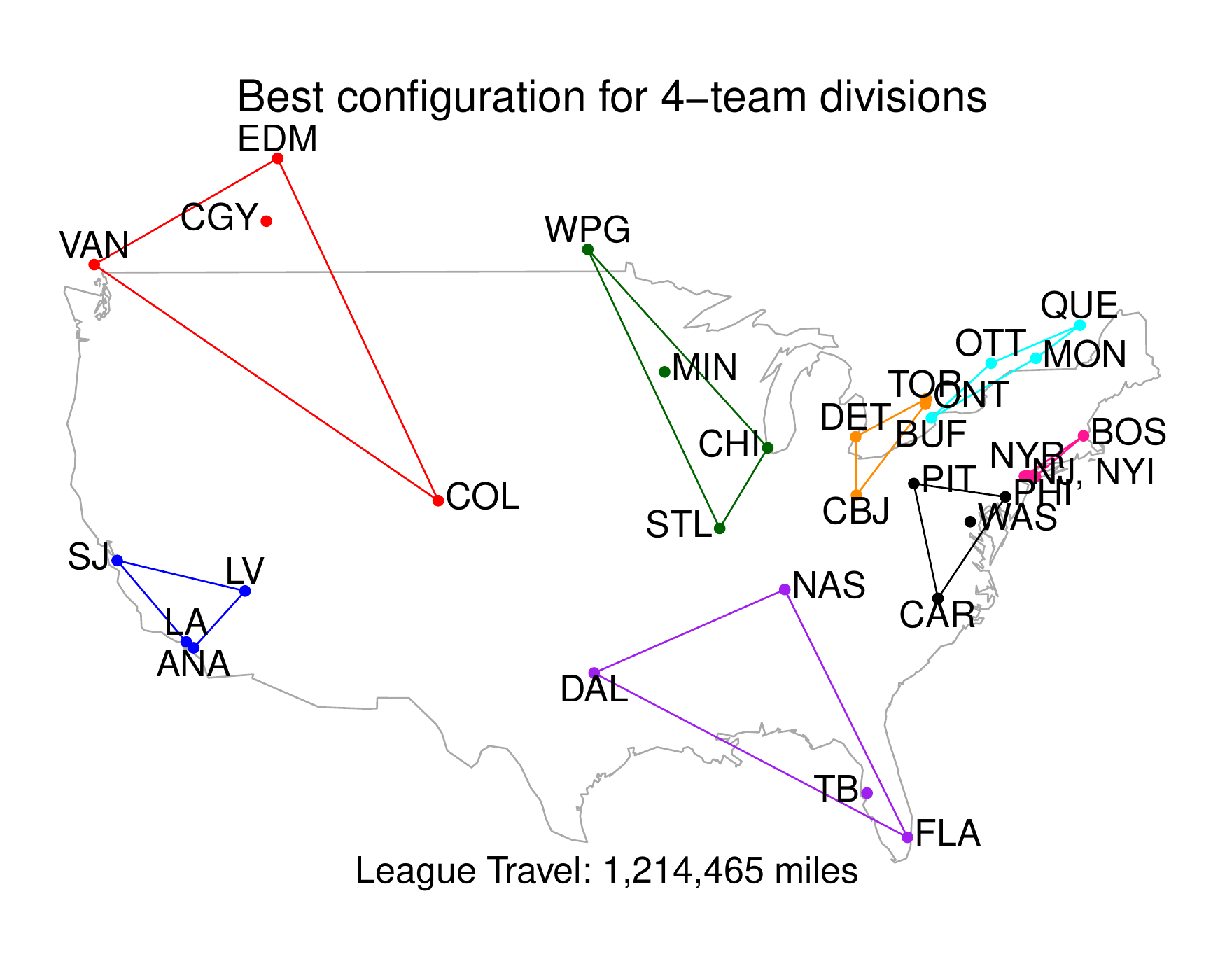}
                \includegraphics[width=.5\textwidth]{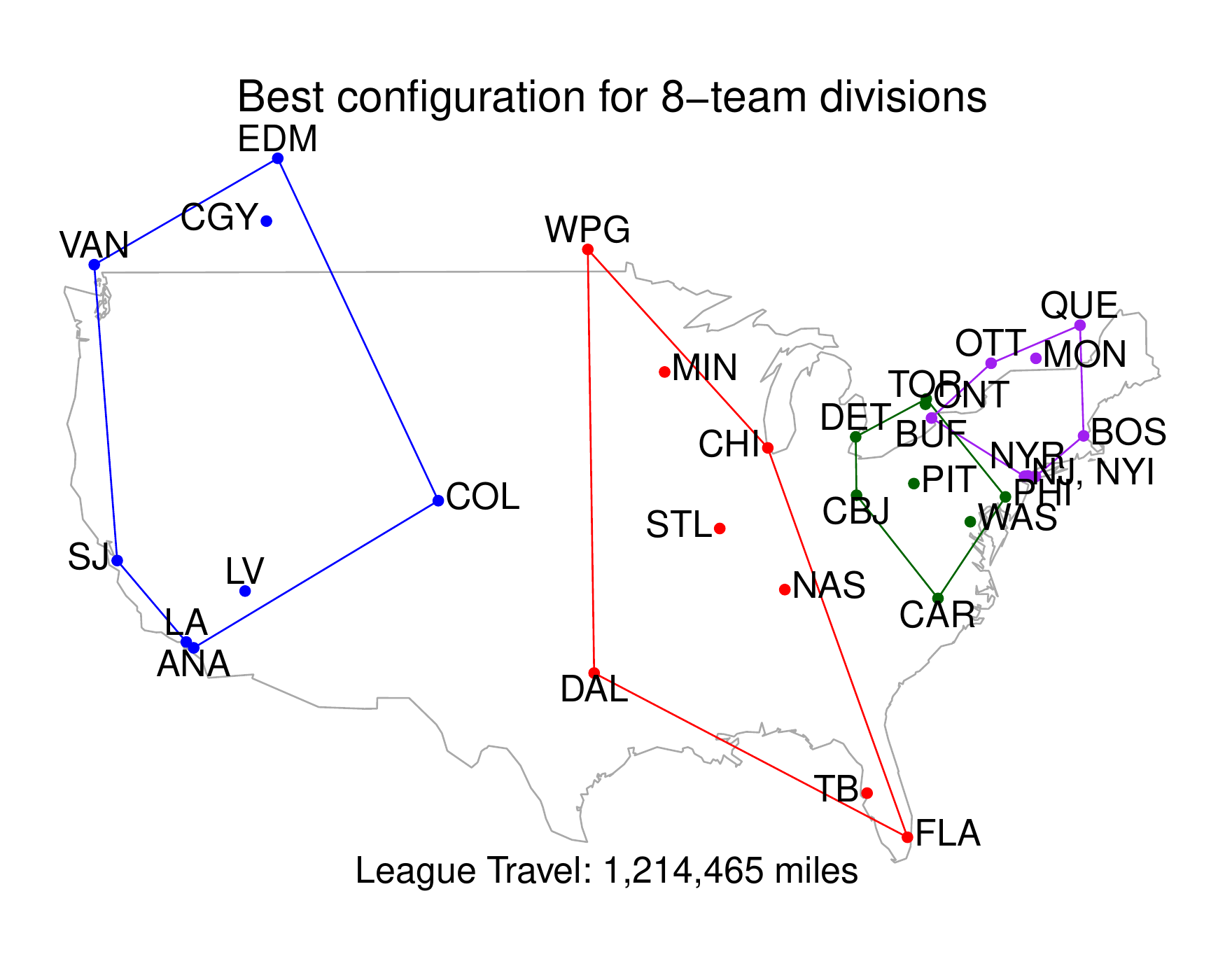}
                \caption{The best solutions if PHO moves to LV and if ONT and QUE are awarded expansion teams. These represent the best solutions assuming $4$-team divisions (left) and $8$-team divisions (right).}
                \label{expansion}
            \end{figure}

        The difference in team travel between using the NHL's proposed $4$-conference configuration and our best $4$-conference configuration is shown in Figure $\ref{proposed-vs-best-4-conf}$.  There are not many big differences in travel, although TB and FLA, two of the teams with the worst travel, would benefit from not being with teams in the northeast.
%
%
%
    \paragraph*{Franchise moves} Our approach can be easily modified to accommodate franchise moves or expansion.   Suppose that, sometime in the near future, the Phoenix Coyotes (PHO) move to Quebec (QUE).  We give the best solution in this case in the top left of Figure $\ref{moves}$%
    , where we have specified the same additional constraints as before.  Note that this solution has TB and FLA as part of the west, so one might prefer to add additional constraint to force TB and FLA to the east. 
    Also, in Figure $\ref{moves}$, we give the best solution if PHO moves to Seattle (SEA), Kansas City (KC), or Houston (HOU).  We note that if PHO moves to southern Ontario, we get the same solution that we got for Quebec, and if PHO moves to Las Vegas, we get the same solutions as if PHO stayed in PHO.  
    \paragraph*{Expansion}
        We can also modify our approach to accommodate for potential future expansion by the NHL.         
        For example, suppose that in a few years PHO moves to LV, and teams in southern Ontario and Quebec are added to the league.  The NHL would have $32$ teams, and would likely choose either four $8$-team divisions or eight $4$-team divisions. 
        We give the best solution under these conditions in the left and right of Figure $\ref{expansion}$, respectively.  
            In the event that the NHL seriously considers expanding to Europe, our approach could be used to estimate league travel miles and the associated cost, as well as propose the best solutions with, for example, one European division of $6$ teams and $5$ North American divisions with $6$ teams each.

\subsection{MLB Realignment}
    MLB has considered different forms of radical realignment over the years.  For example, in $1997$, one very controversial plan involved $4$ divisions of $7$ or $8$ teams each, and the divisions were based on geography so that several teams would have switched from the AL to the NL, and vice versa \cite{realignment-horrible, nytimes-realignment}.  
    MLB has even considered a ``floating realignment'' in which teams could change divisions year-to-year based on things like payroll and a team's plans to contend \cite{floatingrealignment}. 
    
\begin{figure}[h!]
            \includegraphics[width=.5\textwidth]{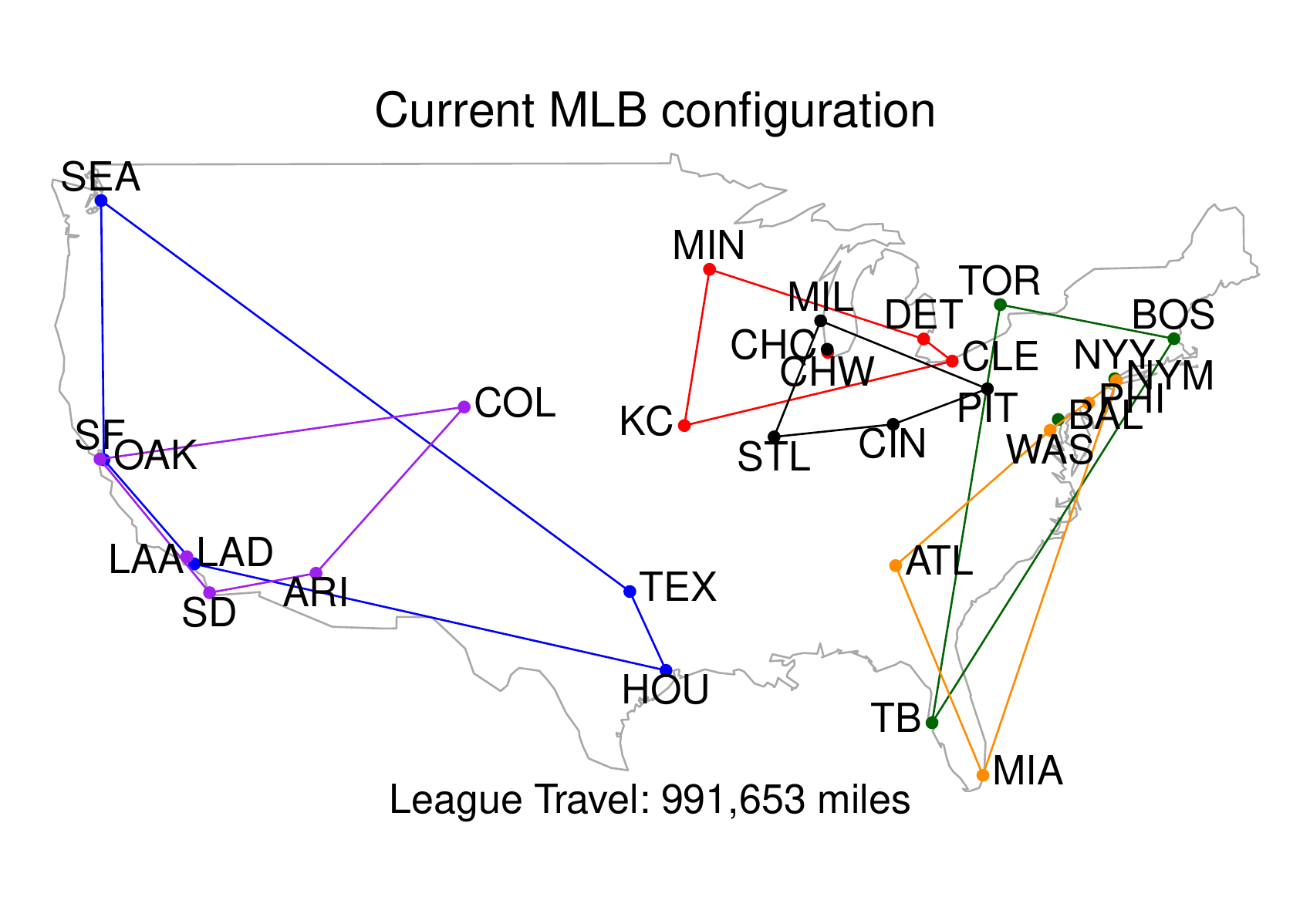}
            \includegraphics[width=.5\textwidth]{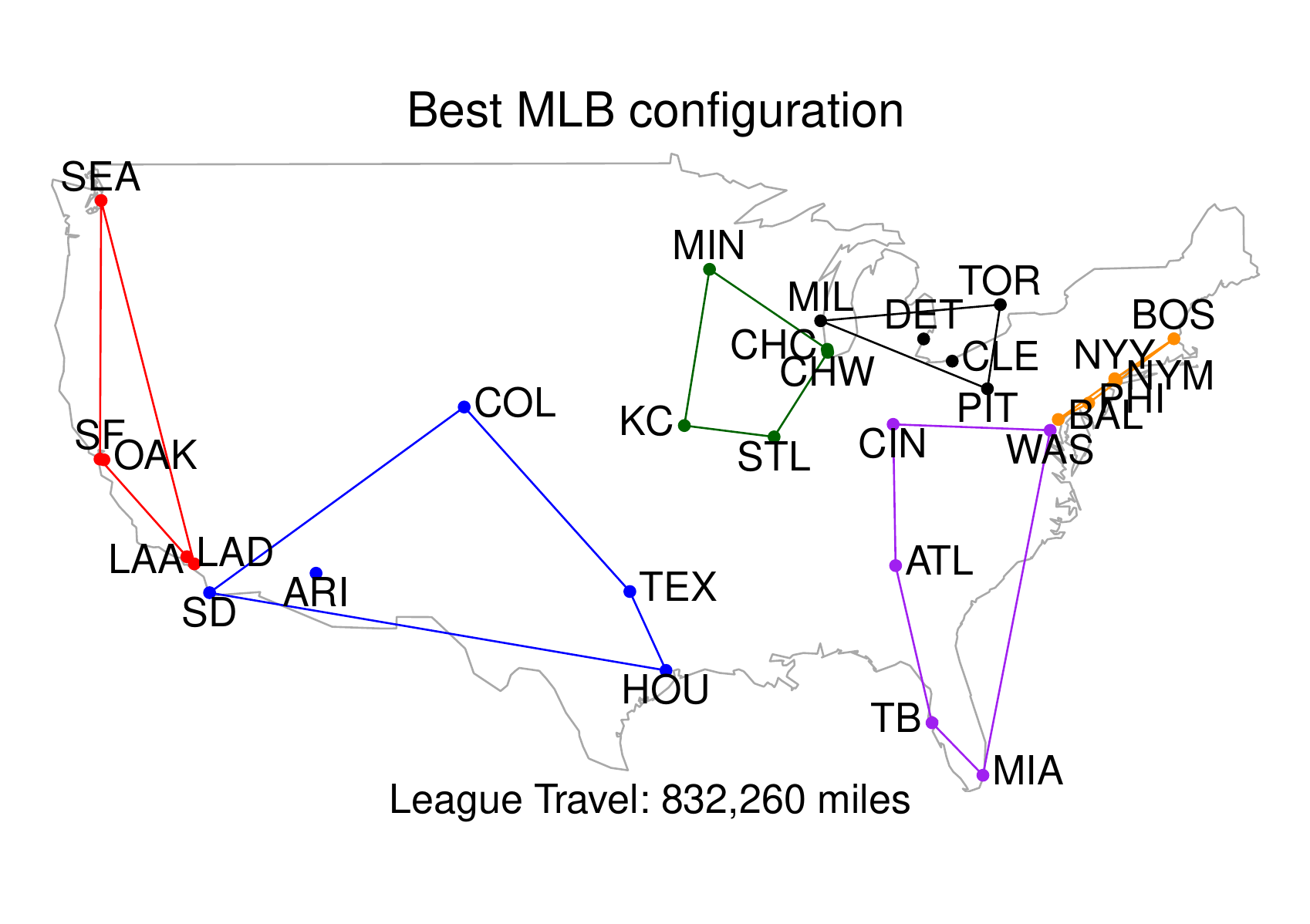}
            \caption{The current MLB alignment (left) and the best MLB alignment (right). 
            }
            \label{mlb}
        \end{figure}
    We give the current alignment and our best $6$-division alignment for MLB in Figure $\ref{mlb}$. Note that we are not requiring that teams stay in their current league.  
    The difference in travel between the current solution and our best solution is substantial.  Current league travel is about $20$\% more than it would be under our best solution.  This is perhaps not terribly surprising, since the pairs of teams in New York, Chicago, and Los Angeles, as well as teams like Philadelphia and Baltimore, Tampa Bay and Miami, San Francisco and Oakland, and Kansas City and St. Louis are not currently in the same division.  

\begin{figure}[h!]
        \centering
            \includegraphics[width=.750\textwidth]{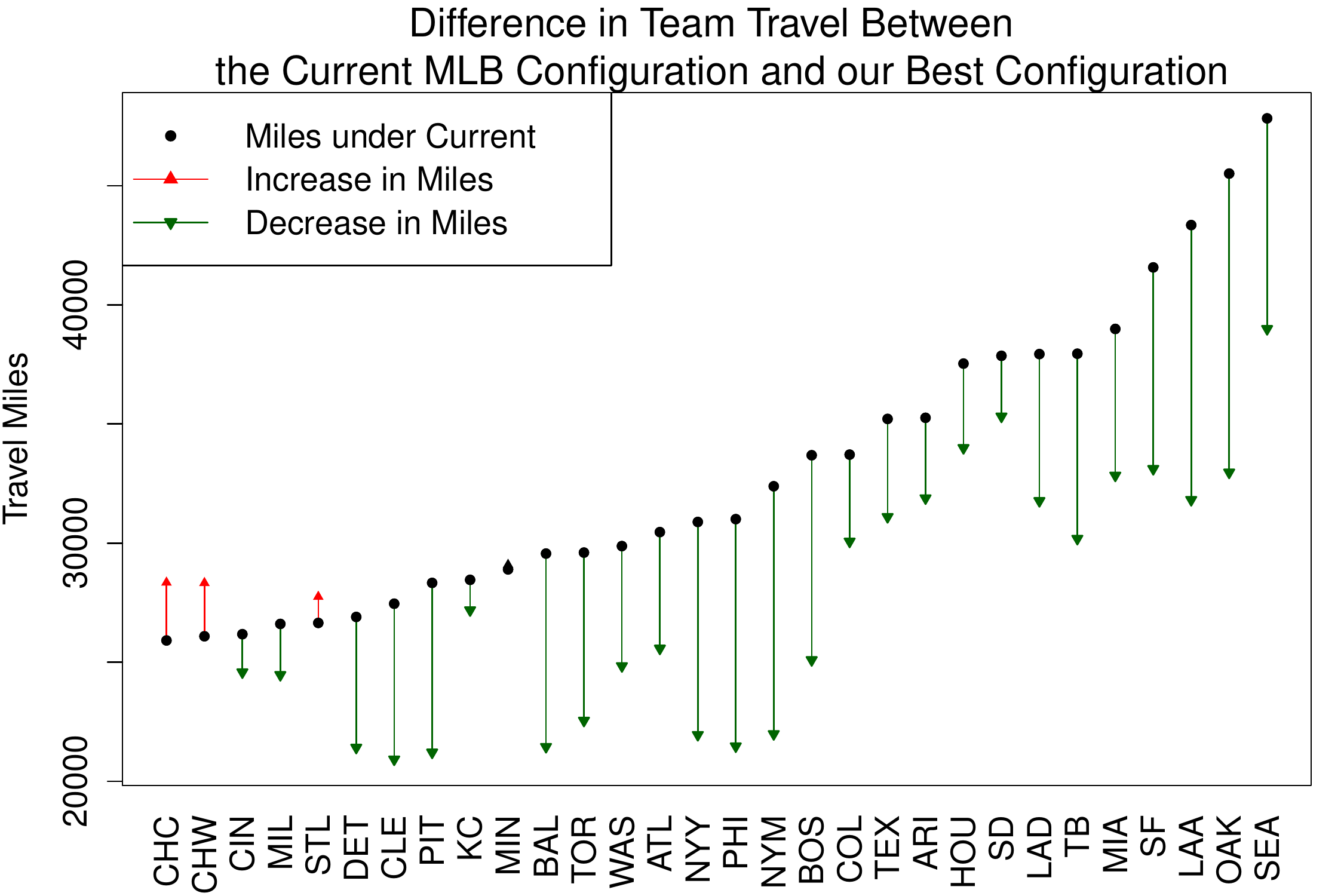}
            \caption{The difference in team travel under the current and best MLB configuration. 
            }  
            \label{current-vs-best-MLB}
        \end{figure}

     We give the difference in team travel under the current MLB configuration and our best configuration in Figure $\ref{current-vs-best-MLB}$. Most teams would have drastically reduced travel, and the teams with the most travel (Seattle, Oakland, LAA, San Francisco, etc.) have among the biggest improvements.  For example, Seattle, the team with the worst travel, would have their travel reduced by $9$,$000$ miles.
     
\begin{figure}[h!]
        \centering
        \includegraphics[width=.49\textwidth]{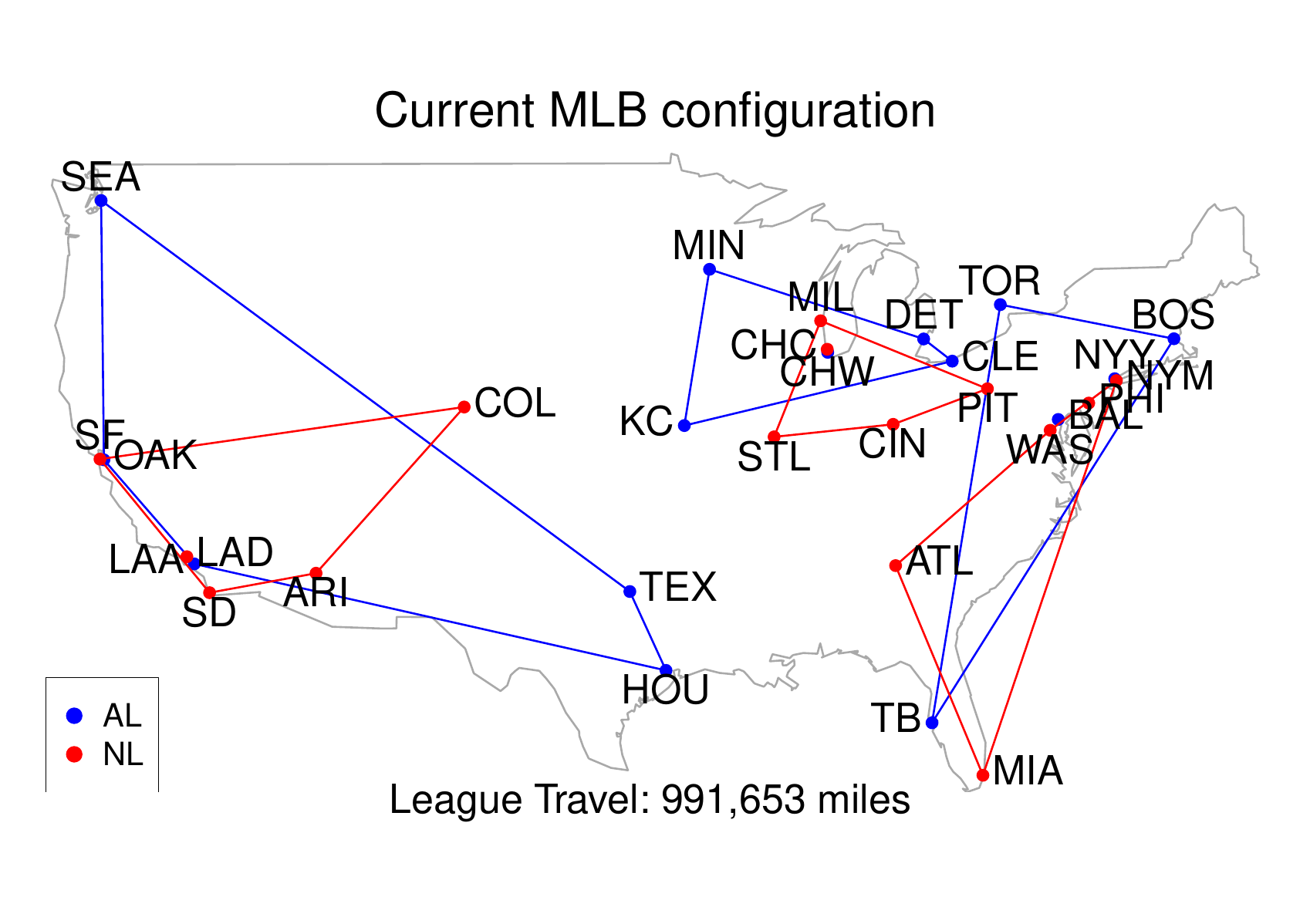}
        \includegraphics[width=.49\textwidth]{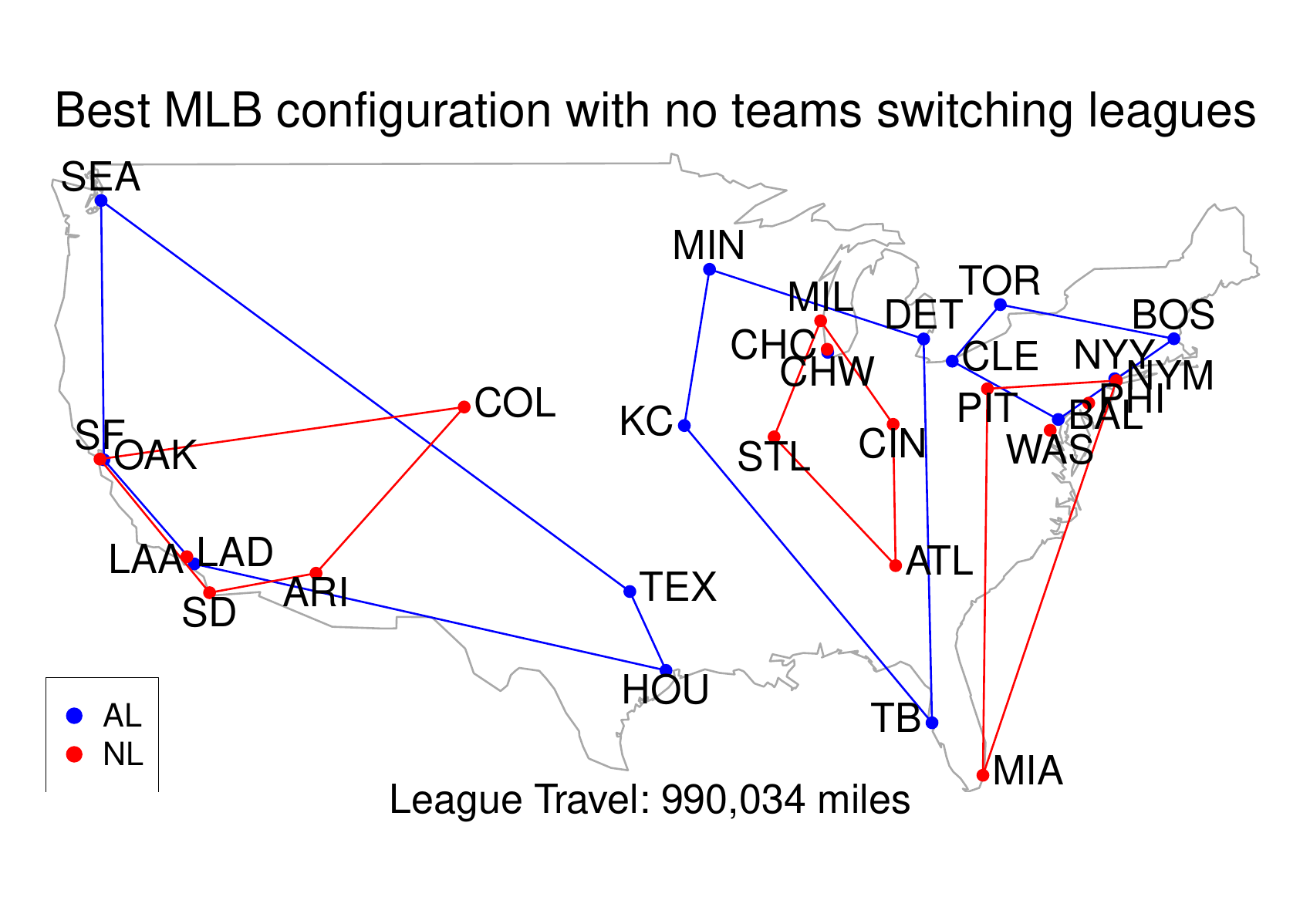}
        \caption{The current MLB configuration (left) and the best MLB configuration in which teams must stay in the same league (right). }  
        \label{fixALNL}
     \end{figure}

MLB might prefer to keep the current American League (AL) and National League (NL) in tact, and would not consider any configuration in which several teams switch leagues as in our best configuration above.  In Figure $\ref{fixALNL}$, we give the best MLB configuration where teams are not allowed to switch leagues (right), along with the current configuration (left).  The NL West and AL West would remain the same in our best solution, and there would only be one change in each league: Atlanta and Pittsburgh would switch places in the NL, and Cleveland and Tampa Bay would switch places in the AL.  This solution would save the league only $2,000$ miles.  

Perhaps some of those opposed to allowing teams to switch leagues would be less opposed after considering the positive impact on the environment our realignment would have.  The current configuration requires roughly $20$\% more travel than our best solution, which means it requires roughly $20$\% more jet fuel.  Suppose teams use a Boeing $747$-$400$, Boeing $747$-$800$, or any other plane that consumes about $5$ gallons per mile \cite{boeing}.  A $160$,$000$ decrease in travel miles during an MLB season corresponds to a decrease of $800$,$000$ gallons of jet fuel per season. 

\subsection{NFL Realignment}
\begin{figure}[h!]
            \includegraphics[width=.5\textwidth]{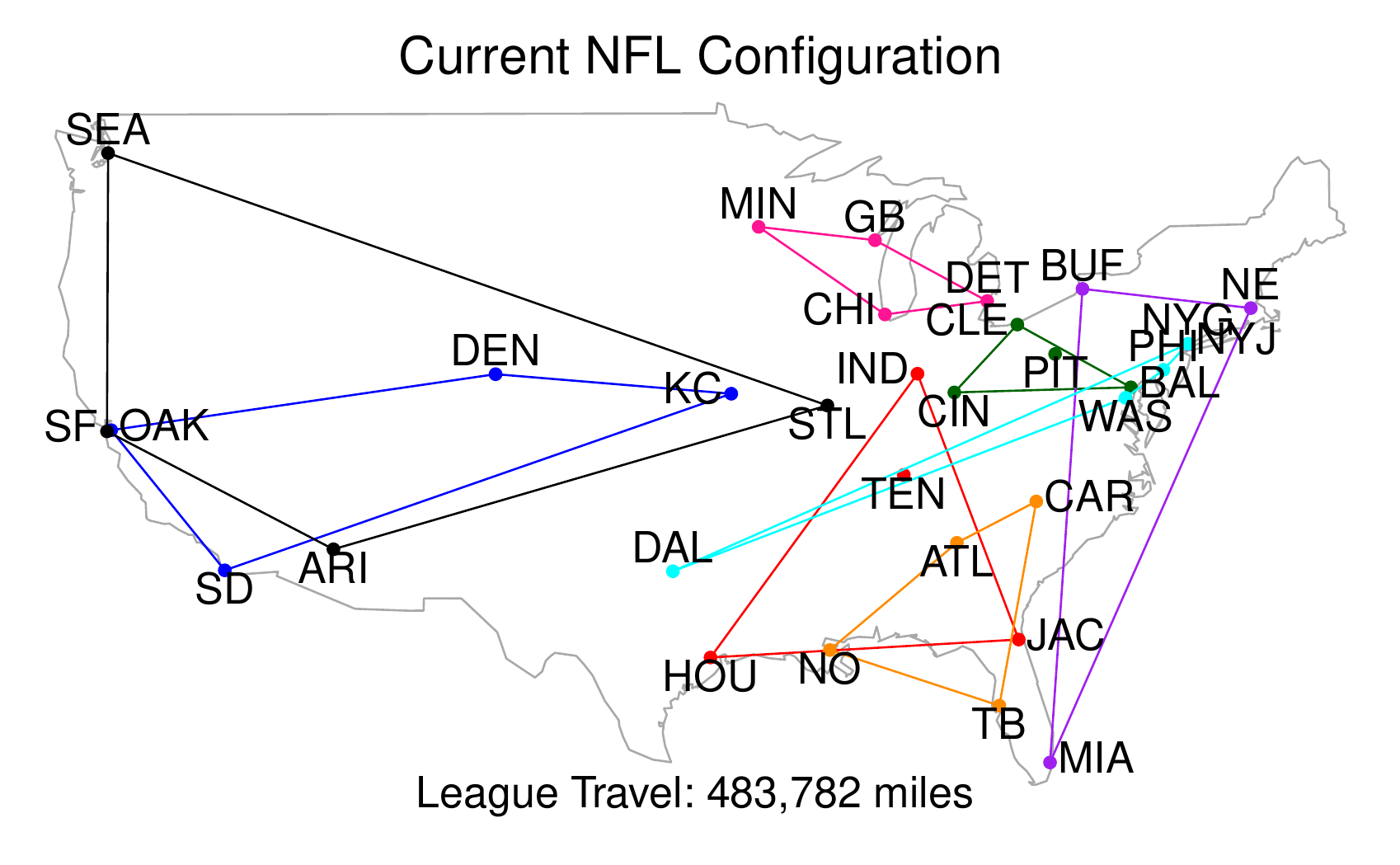}
            \includegraphics[width=.5\textwidth]{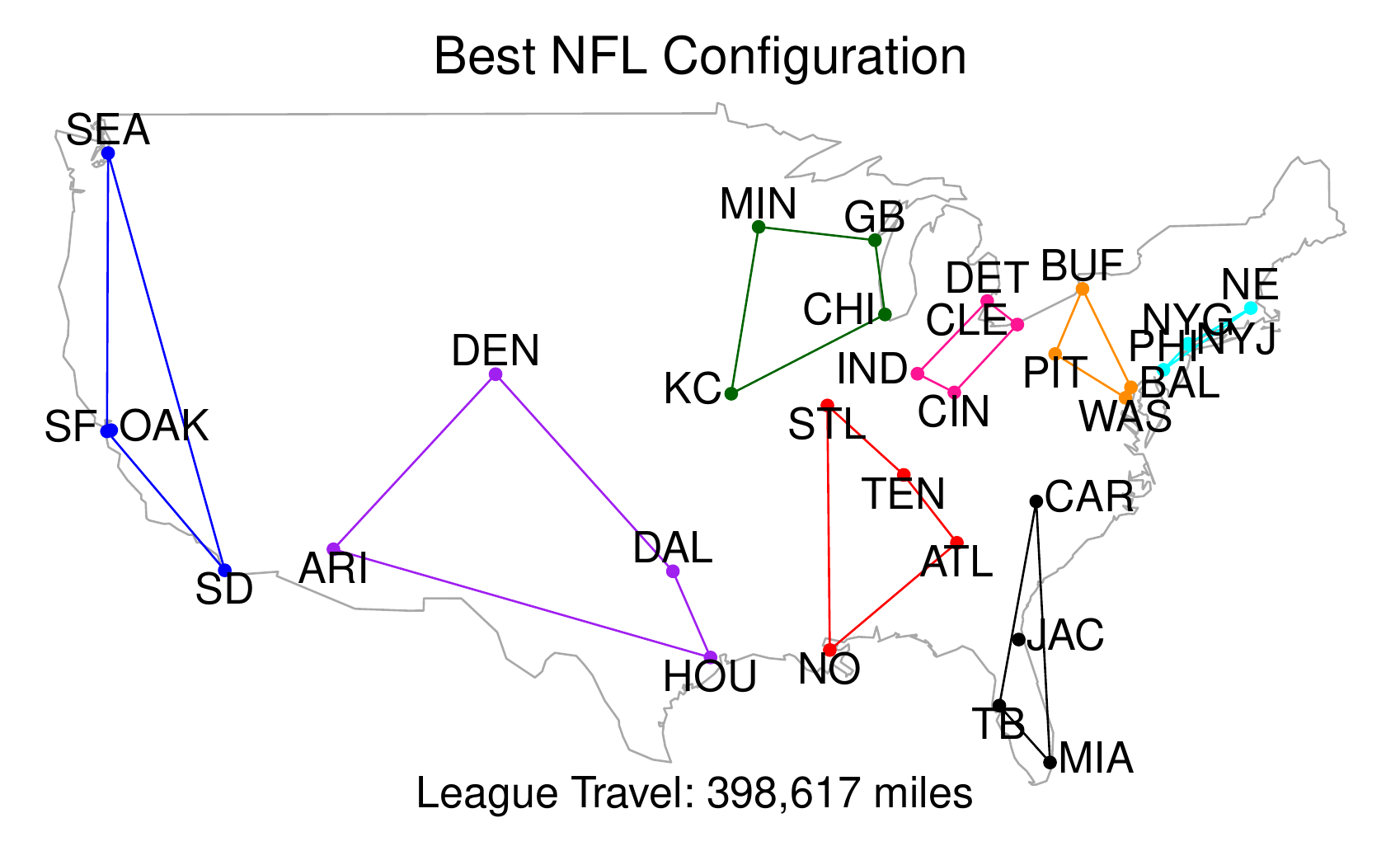}
            \caption{The current NFL alignment (left) and the best NFL alignment (right).  
            }
            \label{nfl}
        \end{figure}

We give the current and best NFL alignment in Figure $\ref{nfl}$.  The current configuration requires $20$\% more travel than the best configuration, but the league still travels far fewer miles than the other three leagues because of their $16$-game schedule.  Still, our best solution would save the league almost $100$,$000$ miles of travel.  

Of course, our best configuration breaks up some rivalries that the NFL might prefer to keep intact.  For example, Dallas is not with their NFC east rivals in our best solution.  One might prefer to add constraints to ensure that these rivalries to stay together.

\begin{figure}[h!]
        \centering
            \includegraphics[width=.90\textwidth]{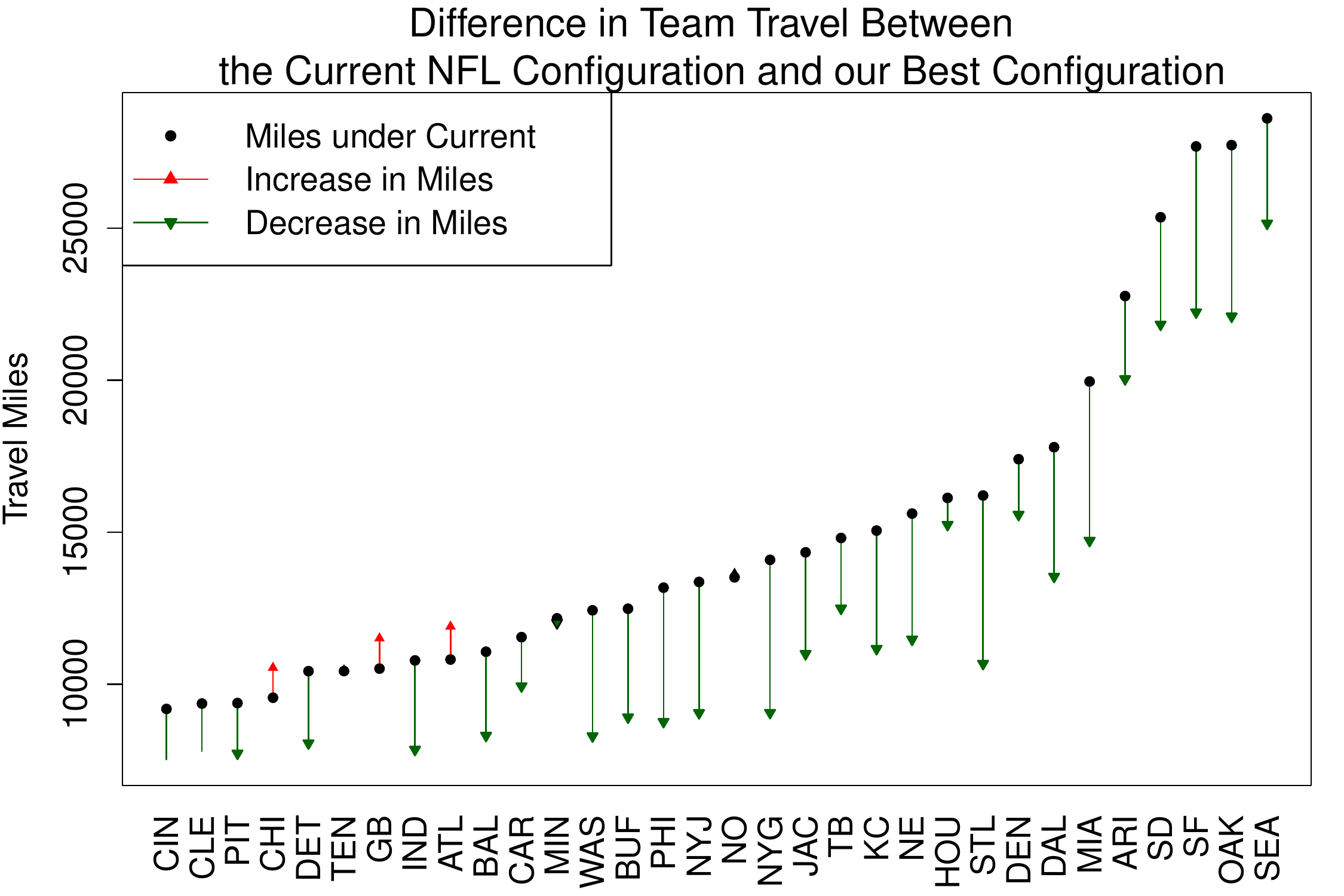}
            \caption{The difference in team travel under the current and best NFL configurations. 
            }
            \label{current-vs-best-NFL}
        \end{figure}

We give the difference in team travel under the current NFL configuration and our best configuration in Figure $\ref{current-vs-best-NFL}$.  Virtually every team would have improved travel in our solution, including all of the teams that have the worst travel.

\subsection{NBA Realignment}
\begin{figure}[h!]
         \includegraphics[width=.5\textwidth]{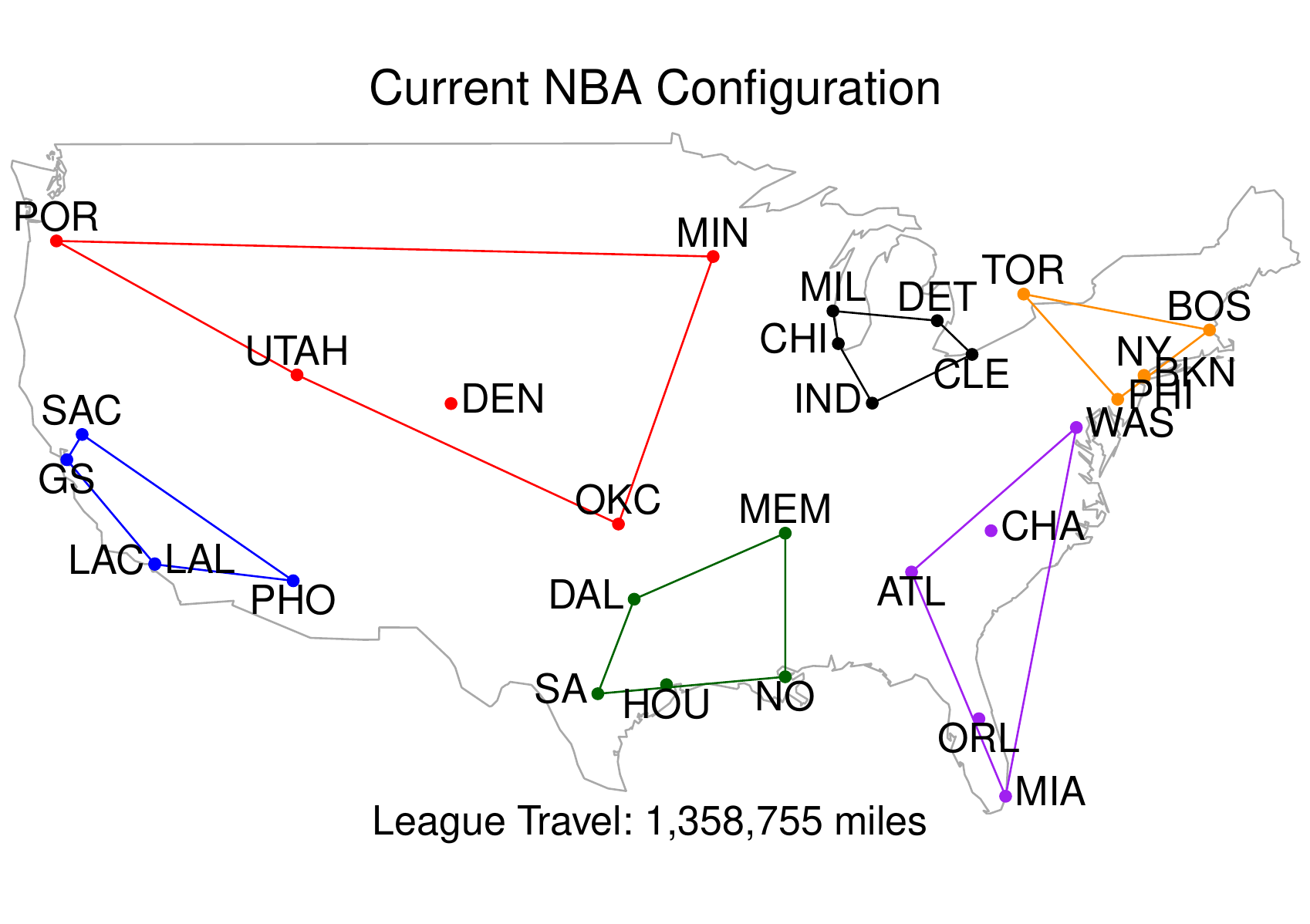}
         \includegraphics[width=.5\textwidth]{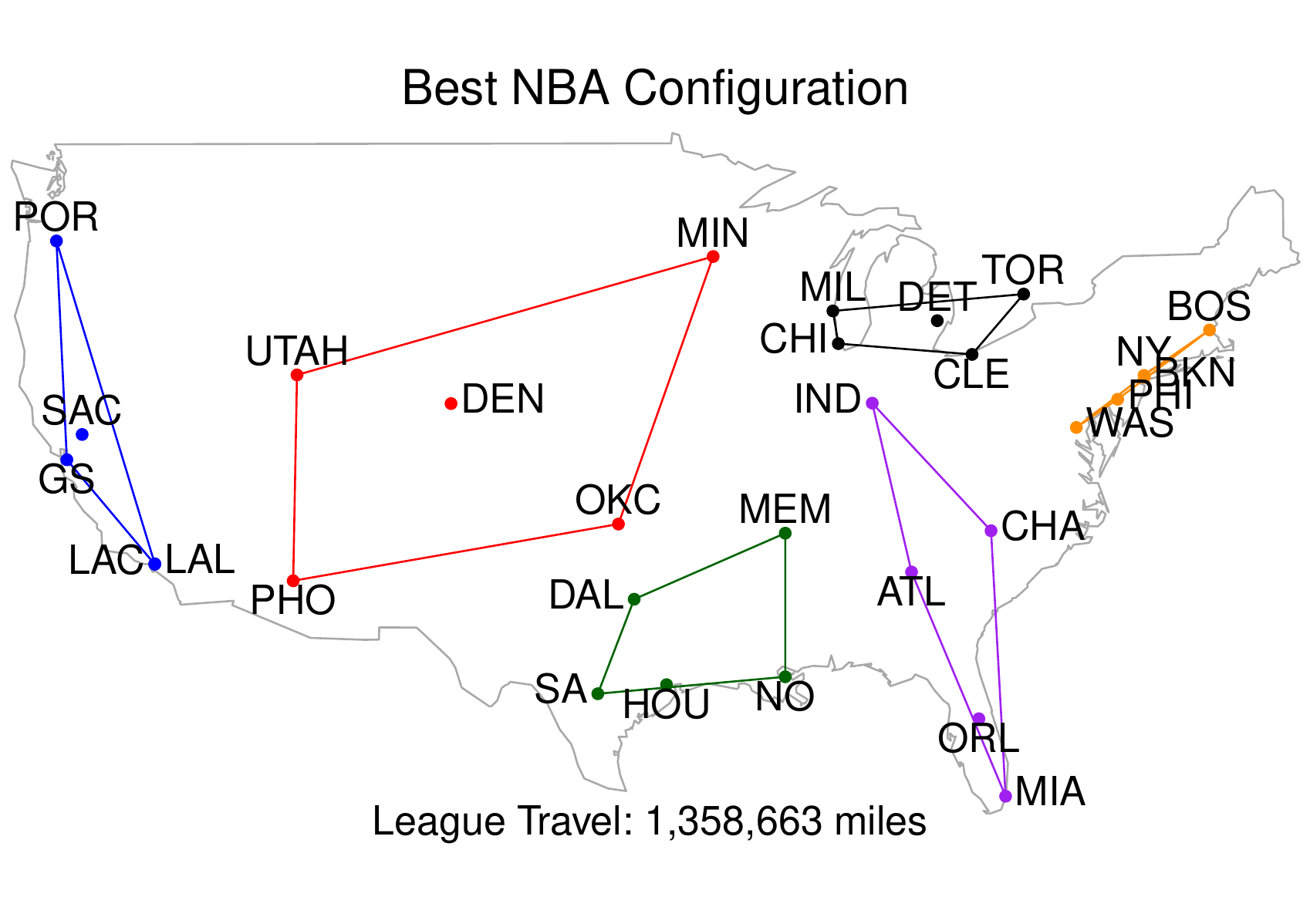}
         \caption{The current NBA alignment (left) and the best NBA alignment (right).  
         }
         \label{nba}
     \end{figure}

    The current and best NBA configurations are given in Figure $\ref{nba}$.      
     Since the league plays a fairly balanced schedule, the current alignment, while not optimal, is only costing the league about $100$ miles in total travel.  Our solution has Portland with the California teams, which probably makes more sense from a time zone standpoint.  The current Northwest division, shown as the red triangle in the upper left, spans three time zones.  After swapping Portland and Phoenix, the division spans only $2$ time zones, and all of the teams in the Pacific division are in the same time zone.   

      \begin{figure}[h!]
      \centering
         \includegraphics[width=.90\textwidth]{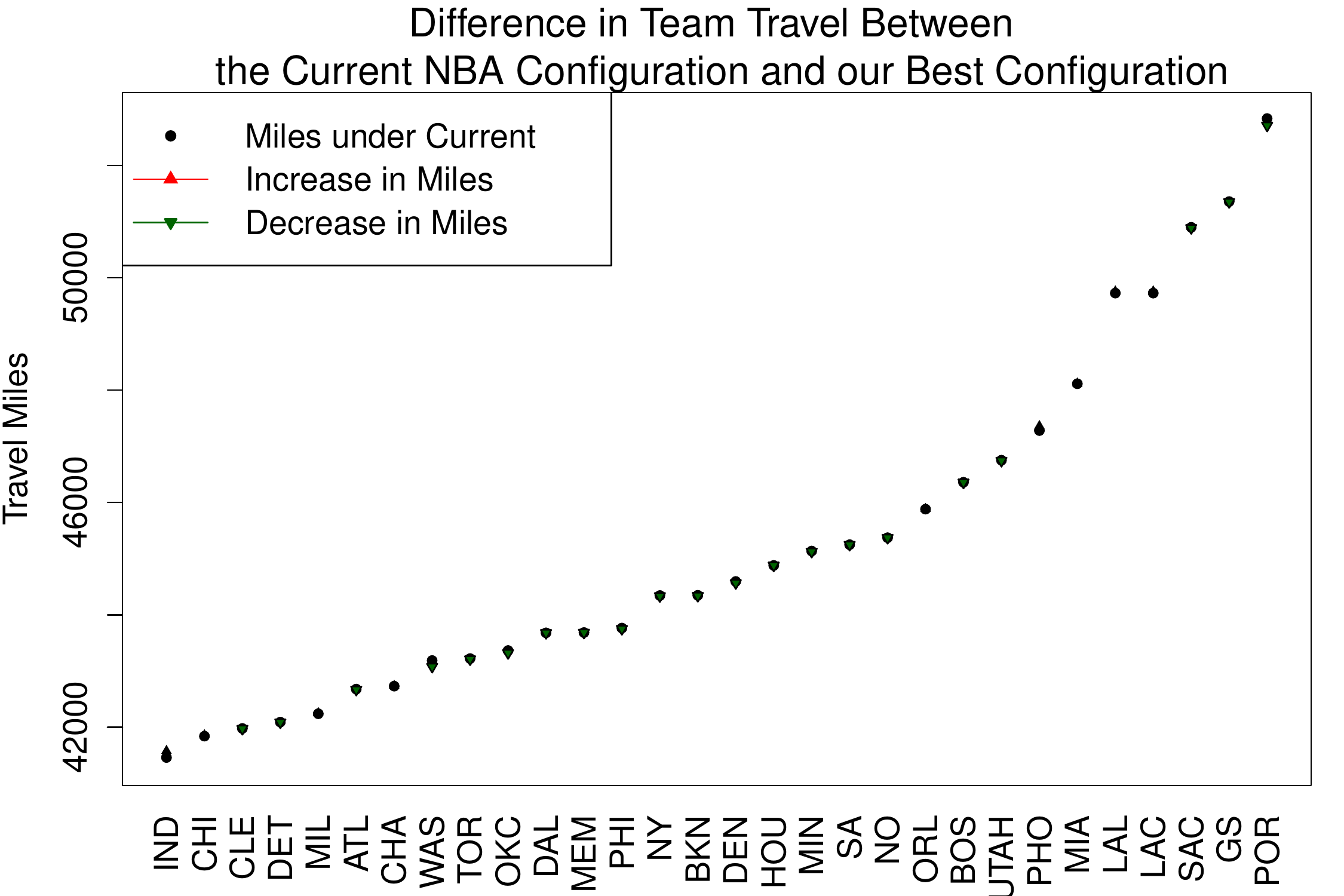}
         \caption{The difference in team travel under the current and best NBA configuration. 
         }
         \label{current-vs-best-NBA}
     \end{figure}         
     We give the difference in team travel under the current NBA configuration and our best configuration in Figure $\ref{current-vs-best-NBA}$.  There are not any major changes in travel, although the team with the worst travel, Portland, has the biggest improvement.  The biggest benefit of our best NBA solution is the time zone improvements mentioned in the previous figure.

\section{Conclusions}
    We have provided a way to estimate team travel in a given league configuration before a schedule is known.  We have also developed a fast way to generate thousands of good solutions for realignment, and we can easily reduce this list by adding any desired constraints.  We can show that the best solution using this method is actually the optimal solution.   We can estimate travel for each of these solutions, but we can also estimate travel for any solution that a league might want to consider, even one we do not generate.  Finally, we have provided a way to visualize any configuration that is under consideration, which could assist humans in making a final decision. 

    In future work, one could attempt to improve the surrogate objective even further by using trends from previous schedules in each of the four leagues.  We could also use our methods in any of the minor leagues associated with the NHL, any of the junior hockey leagues in Canada, any of the hockey leagues in Europe, or any of the minor leagues in baseball.

    Additional information and discussion can be found at \url{www.GreaterThanPlusMinus.com/p/realignment}.  For example, we have posted various animations on that site, including the top $100$ configurations for the NHL, MLB, NFL, and NBA.  We also give animations that help illustrate our algorithm for generating thousands of solutions.

The authors wish to thank Dirk Hoag, Michael Peterson, and Michael Wilczynski for the NHL data, the airplane information and feedback about the paper, and the NBA data, respectively, that were used in this project.

\bibliographystyle{DeGruyter}
\bibliography{realign-bib}

\newpage
\section{Appendix}

\subsection*{Table of Results}

\begin{table}[h!]
\caption{Summary of current and optimized configurations. 
}
\centering
\footnotesize
\begin{tabular}{rrrrrrrrrr}
\toprule
League & 
\multicolumn{1}{m{.6cm}}{\centering \# of Conf} & 
\multicolumn{1}{m{.6cm}}{\centering \# of Divs} & 
\multicolumn{1}{m{.5cm}}{\centering Tms per Div} & 
\multicolumn{1}{m{.5cm}}{\centering Div Gms} & 
\multicolumn{1}{m{.5cm}}{\centering Conf Gms} & 
\multicolumn{1}{m{.5cm}}{\centering non-Conf Gms} & 
Solution &
\multicolumn{1}{m{1.5cm}}{\centering Travel (miles)} &
\multicolumn{1}{m{1.5cm}}{\centering Miles Over Minimum} \\ 
\midrule

\multirow{8}{*}{NHL } 
 & $2$ & $6$  & $5$         & $24$ & $40$         & $18$ &  Best             & $1,155,391$ &   $0$    \\ 
 & $2$ & $6$  & $5$         & $24$ & $40$         & $18$ &  FLA-TB           & $1,155,969$ & $578$ \\ 
 & $2$ & $6$  & $5$         & $24$ & $40$         & $18$ &  Rivalries        & $1,156,530$ & $1,139$ \\ 
 & $2$ & $6$  & $5$         & $24$ & $40$         & $18$ &  $3$ CAN tms    & $1,157,640$ & $2,249$ \\ 
 \vspace{.2cm}
 & $2$ & $6$  & $5$         & $24$ & $40$         & $18$ &  Current          & $1,185,123$ & $29,732$  \\ 

& $4$ & $0$  & $7$, $8$  & $0$   & $36$, $38$ & $46$, $44$ & Best $4$-conf     & $1,228,487$ & $85,437$ \\ 
 & $4$ & $0$  & $7$, $8$  & $0$   & $36$, $38$ & $46$, $44$ & Rivalries      & $1,229,110$ & $86,060$ \\ 
 & $4$ & $0$  & $7$, $8$  & $0$   & $36$, $38$ & $46$, $44$ & Proposed  & $1,245,506$ & $102,456$ \\ 
\midrule
\multirow{3}{*}{MLB } & $2$ & $6$  & $5$         & $24$ & $20$         & $6$ &  Best          & $832,260$ & $0$ \\ 
 & $2$ & $6$  & $5$         & $24$ & $20$         & $6$ &  Current          & $991,653$ & $159,393$ \\ 
 & $2$ & $6$  & $5$         & $24$ & $20$         & $6$ &  Fix AL,NL          & $990,034$ & $157,774$ \\ 

\midrule
\multirow{2}{*}{NFL } & $2$ & $8$  & $4$         & $2$ & $0$,$1$         & $0$,$1$ &  Best         & $398,617$ &   $0$    \\ 
 & $2$ & $8$  & $4$         & $2$ & $0$,$1$         & $0$,$1$ &  Current         & $483,782$ &   $85,165$     \\ 

\midrule

\multirow{2}{*}{NBA } & $2$ & $6$  & $5$         & $16$ & $36$         & $30$ &  Best         & $1,358,663$ &   $0$   \\ 
 & $2$ & $6$  & $5$         & $16$ & $36$         & $30$ &  Current         & $1,358,755$ &   $92$  \\ 

    \bottomrule
\end{tabular}  
\label{summary-table}
\end{table}

\paragraph{Minimizing Total League Travel with Mixed Integer Programming }
        
We now outline a way to find the provably optimal solution for each league.  We note that this is a similar problem to that studied in \cite{mitchell}, but we aim to minimize total league travel distances as opposed to minimizing intradivisional travel distance.  

The problem of finding a league structure for which the surrogate objective is minimized can be formulated as an Integer Programming Problem.
We have a set $T$ of $n$ teams/cities and a set $S$ of $s$ divisions.
For any two teams $u,v \in T$ recall that $d(u,v) = d(v,u)$ is the travel distance between the home cities of  $u$ and $v$. 

The input data consists of the following: 
\begin{itemize}
\item{$D =( d(u,v): u,v \in T)$ is the  $n \times n$ inter-city distance matrix.}
\item{ $G $ is the $s \times s$ away game matrix.}
For each pair $(i,j)$ of divisions, $G_{ij}$ specifies the number of away games to be played by teams in division $i$ against teams in division $j$.  In the case that this number is not the same for all pairs of teams in these divisions, we set $G_{i}j $ equal to the average number of games over pairs  of teams in the two divisions.  When considering intra-divisional games, that is, when $i = j$, we only consider pairs of distinct teams. 

\item{ Let $d$ be an $s$ element vector  with $d_i$ equal to the number of teams required in division $i$.}  Note that  $\sum_{i} d_i = n.$ 
\end{itemize}	
In the case of the current NHL, $n = 30$, $s = 6$,   $d = [\begin{array}{c c c c c c}  5& 5& 5& 5 &5& 5 \end{array}]$.  The away game matrix 
 
$$G = \left[ \begin {array}{c c c c c c}
	3 & 2 & 2 & .6 & .6 & .6 \\ 
	2 & 3 & 2 & .6 & .6 & .6 \\
	2 & 2 & 3 & .6 & .6 & .6 \\
	.6 & .6 & .6 & 3 & 2 & 2 \\
	.6 & .6 & .6 & 2 & 3 & 2 \\
	.6 & .6 & .6 & 2 & 2 & 3 \end{array}  \right] .$$

For each team $v$ and each division $i$ we have a variable $x_{vi} = 1$ if team $i$ is in division $i$ and $x_{vi} = 0$ if not.

In order to  evaluate the quadratic objective function, we  define a set of variables $y_{uvij}$ as follows:  For each pair $(u,v)$ of teams and for each pair $(i,j)$ of divisions, we have 
$y_{uvij}=1$ if team $u$ is assigned to division $i$ and team $v$ is assigned to conference $j$ and 
$y_{uvij}=0$ if not.  The cost $c_{uvij}$ of $y_{uvij}$ is defined to be $c_{uvij} = D_{uv} \cdot G_{ij}$.
(This enables us to correctly evaluate the quadratic objective function.)

In our example, we have $30 \times 6 = 180$ variables $x_{vi}$ and $180^2 = 32,400$ variables $y_{uvij}$.

There are three sets of constraints on our variable.  The first set ensures that we have the correct number of teams in each division and that each team belongs to a division:

$$ \sum_{v \in T} x_{vi} = d_i \mbox{ for each division }  i; $$ 

$$ \sum_{i \in S} x_{vi} = 1 \mbox{ for each team }  v. $$

The second ensures that each pair of teams play in exactly one pair of conferences.  

$$\mbox{For each pair $u,v$  of cities, } \sum_{i,j \in S} y_{uvij}=1; $$
$$\mbox{For each pair $i,j$  of divisions, } \sum_{u,v \in T} y_{uvij}=d_i \cdot d_j. $$

The third set of constraints forces the $x$ and $y$ variables to behave consistently.  We want to have $y_{uvij} = 1$ only if $x_{ui} = 1$ and $x_{vj}=1$ and equal $y_{uvij} = 0$ otherwise.  We create the inequalities 
\begin{equation}  y_{uvij} \leq 0.5(x_{ui} + x_{vj}) \mbox{ for all } u,v \in T, i,j \in S.\label{link} \end{equation}

We also constrain $y_{uvij}$ to be a $0-1$ variable for all $u,v,i,j$.  This forces $y_{uvij}$ to be $0$ unless both $x_{ui} = 1$ and $x_{vj}=1$. Finally, we define a new  cost function $c^\prime$  by letting 
$c^\prime_{uvij} = M - c_{uvij}$ for all $u,v \in T$ and $i,j \in S$, where $M$ is a constant larger than any cost $c_{uvij}$.  Then minimizing $c$ is equivalent to maximizing $c^\prime$ and all $c^\prime_{uvij}$ are strictly positive.

Each variable $y_{uvij}$  occurs in a single inequality (\ref{link}) so if we maximize the objective, every $y$ will take on the value  $1$ if and only if both $x_{ui} = 1$ and $x_{vj}=1$, as we desired.  So, finally, the mixed integer programming problem that we solve to obtain an optimal league structure is 

$$\mbox{ maximize} \sum_{u,v \in T, i,j \in S} y_{uvij} \cdot c^\prime_{uvij} $$

subject to 

$$ \sum_{v \in T} x_{vi} = d_i \mbox{ for each division }  i; $$ 

$$ \sum_{i \in S} x_{vi} = 1 \mbox{ for each team }  v; $$

$$ \sum_{i,j \in S} y_{uvij}=1 \mbox{ for each pair $u,v$  of cities }; $$
$$ \sum_{u,v \in T} y_{uvij}=d_i \cdot d_j \mbox{ for each pair $i,j$  of divisions}; $$
$$  y_{uvij} \leq 0.5(x_{ui} + x_{vj})  \mbox{ for all } u,v \in T, i,j \in S; $$
$$ x_{uv} \geq 0  \mbox{ for all } u,v \in T \mbox{  and } y_{ij} \geq 0, \mbox{ integer for all } i,j\in S.$$

It is straightforward to add extra constraints to this model to require certain teams, or combinations of teams to be in specified divisions. We also were able to significantly improve performance by providing CPLEX with a starting solution equal to the best solution found by our heuristic for the problem.  Also, constraining pairs of cities on opposite sides of the continent to be in different divisions significantly improved solution time.  

The solution time required to solve these league structure problems ranged from several hours to tens of hours on a moderately powerful workstation.  However, as noted earlier, this only produced a single, provably optimal, structure.  The set of optimal and near optimal solutions provided by the heuristic provide more options to league planners.

\end{document}